\documentclass[10pt,a4paper]{article}
\usepackage[latin1]{inputenc}
\usepackage{amsmath}
\usepackage{amsfonts}
\usepackage{amssymb}
\usepackage{graphicx}
\usepackage{natbib}
\usepackage{tikz}
\usepackage[colorlinks=true,urlcolor=blue]{hyperref}
\usepackage{multirow}
\usetikzlibrary{patterns}
\usetikzlibrary{decorations.pathmorphing}
\usepackage{algorithm}
\usepackage{algorithmic}

\begin{document}
\title{Multi-team Formation using Community Based Approach in Real-World Networks\footnote{This work of Ramesh is jointly sponsored by AICTE,INDIA and BVRIT, Narsapur, Telangana}}

\date{}
\author{Ramesh Bobby Addanki\footnote{ramesh.ab@bvrit.ac.in, abrameshba@gmail.com} \\ {\small SCIS, University of Hyderabad}\and S. Durga Bhavani\footnote{sdbcs@uohyd.ernet.in}\\{\small SCIS, University of Hyderabad}}

\maketitle

\begin{abstract}
	
	In an organization,  tasks called projects that require several skills, are generally assigned to  teams rather than  individuals. 
	The problem of choosing a right team for a given task with minimal communication cost is known as {\it team formation problem} and many algorithms have been proposed in the literature. 
	We propose an algorithm that exploits the community structure of the social network and forms a team by choosing a leader along with its neighbours from within a community. This algorithm is different from the skill-centric algorithms in the literature which start by searching for  each skill, the suitable experts and do not explicitly consider the structure of the underlying social network. The strategy of community-based team formation called TFC leads to a scalable approach that obtains teams within reasonable time over very large networks. Further, for one task our algorithms {\it TFC-R} and {\it TFC-N} generate multiple teams from the communities which is  show-cased as a case-study in the paper. 
	
	The experimentation is carried out on the well-known DBLP data set where the task is considered as writing a research paper and the words of the title are considered as skills.  Team formation problem is translated to finding possible authors for the given paper, who have the required skills and having  least communication cost.   In the process, we build a much larger bench-mark data set from DBLP for team formation for experimentation. We do not retrieve communities using community discovery algorithms, but consider the subsets of DBLP based on research areas like DB and  VLDB as communities. Clearly there is a trade-off between the time taken and communication cost. Even though the benchmark algorithm {\it Rarestfirst}  takes least time, our algorithms {\it TFC-N} and {\it TFC-R} give much better communication cost. They also  outperform the standard algorithms  like {\it MinLD} and {\it MinSD}  with respect to the time taken in finding a team. Further, the teams found by {\it TFC-N} show similar or lesser communication cost in comparison. The time taken by our algorithms on communities are several orders faster than the time taken on the larger network without compromising too much on the communication cost. \\
	\textbf{keywords :\textbf{\textbf{}}} 	social networks, team formation, degree centrality, power law, DBLP network, communities 
\end{abstract}
\section{Introduction}

	Many real-world challenges like hackathons,  community-based software development or the software challenges thrown open by major conferences are examples of tasks requiring multiple skills that can only be tackled as a team. 
	Teams in the conventional sense are co-located and consist of individuals working in physical proximity. On the other hand, in the context of software industry or even technical hackathons, typically the teams are built across geographical boundaries. Hence it is important to find algorithms that can constitute teams having the skills required as well as  keep the cost of the project within the budget. Conventionally, members have been largely selected based on their functional skills. There may or may not be collaboration among team members. On the other hand, many studies like \cite{Joseph-2020,Marr-2016} show that insufficient communication and ineffective management are some of the reasons for project failure. 	As per \cite{Tavrizyan-2019} only 2.5\% of the companies complete their projects 100\% successfully.  
	Josh Steimle \cite{Josh-2019}, claim that most of the technology implementation projects fail because of people, and  not due to technology. \cite{Lappas-finding-2009} were the first to incorporate this aspect into the mathematical model of the team formation problem by embedding the members on a social network and including communication cost as one of the main costs of the project.

	
	In the era of information and communication technology, there is an explosion of social networking sites that have emerged in the last decade  providing a platform to share common interests. 	The volume of interactions has increased tremendously leading to very large and modular networks. Hence there is a need for algorithms that are scalable for  social networks.
	
	In this paper we propose an algorithm that takes advantage of the power law and the  community structure that is intrinsic to a social network. Our algorithm adopts a kind of divide and conquer strategy. First a community of members having majority of the skills required for the task is retrieved from the social network. Then a leader is picked from the heavy tail of the community who in turn chooses team members having expertise from within his/her neighbourhood. 
	This certainly ensures good communication among the team members.  Since this algorithm works at community-level rather than the whole network, it runs very fast, much faster than many of the well-established algorithms in the literature. 
	
	The organization of the paper is as follows. The problem statement and the notation required is set in Section~\ref{sec:background} and the related literature is given in Section~\ref{sec:rel-lit}. The algorithms of TFC-R and TFC-N are proposed in Section~\ref{sec:algo}. The construction of the dataset is described in Section~\ref{sec:benchmark}. In Section~\ref{sec:network-analyis}, the insights gained from  network analysis of the DBLP data set with respect to the skill coverage vs expert degree distribution are explained. Finally in Sections~\ref{sec:results} and \ref{sec:results-comm}, the comparative results obtained for the algorithms in terms of execution time and the various communication cost measures and the community-wise performance  of the algorithm are given. The paper ends by discussing the results of the algorithm in terms of the multiple teams obtained for each task on the case-study of \cite{Wang-comparative-2015} in Section~\ref{case-study} followed by conclusions.


	\section{Background}\label{sec:background}
	A social network is modeled as a graph $G = (V, E)$ in which experts are considered as nodes and their mutual interactions are depicted as edges with weights given by a distance measure based on the strength/weakness of the interaction. 	

	\subsection{Notation}\label{ssec:notation}
	The notation required for the team formation problem formulation is given in Table~\ref{tab:notation}.

\begin{table}[!h]
	\centering
	\begin{tabular}{|l|l|}
		\hline
		Symbol& Description\\
		\hline
		$G$&Graph representation of network\\
		\hline
		$V$&Set of experts as nodes for $G$\\
		\hline
		$E$&Set of weighted edges for $G$\\
		\hline
		$S$& Universal set of skills of $V$\\
		\hline
		$T$& Set of skills required for a task $T$\\
		\hline
		$s(v)$&Set of skills possessed by an expert $v \in V$\\
		\hline
		\multirow{2}{*}{$HD$}& Set of high degree nodes  \\
		&having at least one skill in T \\
		\hline
	\end{tabular}
	\caption{Notation used for the Team formation problem}
	\label{tab:notation}
\end{table}

	\subsection{Problem statement}
	Given a set of experts $V$ in $G$ and a set of skills $S$, an expert $v\in V$ is associated with skill set represented as $ s(v) $ and given a task $T =\{ s_{1}, s_{2}, ..., s_{k} \}$  contained in $S$, {\it team formation} problem is to find a team  $X\subseteq V$  such that $ \bigcup_{v \in X}^{}s(v) \supseteq T $, such that the communication cost of the team $X$ is minimized.

	\subsection{Communication cost}\label{sec:comm-cost}
	A few of the popular  communication cost functions that measure the quality of a team formation algorithm(TF) as given in a survey of \citet{Wang-comparative-2015} are described below.

	Let $sp(u, v)$ denote the weight of the shortest path found between nodes $u$ and $v$ in the graph $G$. 
	\begin{itemize}
	\item {\it Diameter }: Diameter of a team $X$ is defined as the weight of the longest among the shortest paths between all pairs of nodes of the team $X$.
	\item {\it Sum Distance} :  A measure that computes the sum of distances between each pair of skills (i.e. experts chosen by the algorithm for the skill) of task $T$.
	\begin{equation}
	\sum_{i=1}^{|T|}\sum_{j=i+1}^{|T|}sp(v_{i}, v_{j})
	\end{equation}
	where $v_{i}, v_{j}$ are experts responsible for skills $i, j$ in $T$ and $v_{i}, v_{j} \in X$. Since one expert may be responsible for more than one skill, the cost is calculated with respect skills rather than experts.
	\item {\it Leader Distance } : Sum of distances from leader $v_L$ in the team to other members of the team.
	\begin{equation}
	\sum_{v_{i}\in X, v_{i}\ne v_{L}}sp(v_{i}, v_{L})
	\end{equation}
	where $v_{L}$ is leader and $v_{i}, v_{L} \in X$.
	\end{itemize}

	\section{Related literature}\label{sec:rel-lit}

	In the field of combinatorial optimization one of the classical problems is assignment problem also referred to as  task assignment problem(TAP). 
	In TAP, given $ m $  agents with certain skills and $ n $ tasks, the problem is to find an assignment that matches the tasks with the agents having the required skills. 
	Since each agent incurs a cost, the problem is that of finding an assignment of minimum cost. 
	With a constraint that at most one agent can be assigned to each task and at most one task to each agent, TAP is to find an allocation that maximizes task allocation with minimum cost. 
	\Citet{Lappas-finding-2009} were the first to introduce social network into the assignment problem. They name the agents as \textquoteleft experts' and  that the experts are interacting on a social network. They propose the problem of  Team formation (TFP) as one in which given a task requiring a set of skills, the problem is to find a team of experts who can perform the task incurring minimum communication cost. As explained in \ref{sec:comm-cost}, there are many communication costs that have been proposed in the literature. \citet{Lappas-finding-2009} propose two communication cost measures for evaluating the collaboration among the team members, namely diameter and Steiner tree communication costs.  \citet{Gaston-complex-2003} note that teams from scale-free networks perform well.

	\citet{McDonald-recommend-2003} observe that workplace collaborations are strongly influenced by social relationships. 
	In the same year \citet{Wi-knowledge-2009} proposed an algorithm based on finding a  team leader who then identifies the  team members. They used a multi-objective fuzzy model giving importance to both interpersonal (collaboration) and technical skills(knowledge) for team formation. 

	Several variations of TFP have been proposed in the literature.   \citet{Li-generalized-2010,Amita-multi-2012} consider  redundant number of experts demanded for each skill.   \citet{Aris-online-2012}, constrain the problem by allowing   each agent to participate in more than one project with an upper bound called as workload of the agent. The authors propose two greedy heuristics in this work.
	An online team formation problem is proposed by \citet{Aris-power-2010,Aris-online-2012}, in which formation of multiple teams is considered where the  tasks keep coming online. The experts that have been assigned for a team may not be available for the next task and hence the algorithm needs to keep track of the workload of the agents. This problem is termed as Balanced social task assignment problem(BSTAP). 

	Limiting maximum skills contributed by an expert to a project is called the  \textit{capacitated team formation problem} \cite{Majumder-capacited-2012}. 
	Problem defined by \citet{Lappas-finding-2009} is further extended by \citet{Kargar-efficient-2012} by adding financial cost of agents to the team cost. They proposed heuristics for this bi-objecive TFP that optimizes both financial cost and communication cost. 
	TFP has been further expanded by \citet{Kargar-discovering-2011} who propose two new communication cost measures to evaluate the team, namely, leader distance and sum distance that have been defined in Section~\ref{sec:comm-cost}. 

	We find that in the  literature, TFP has not been limited to skill coverage and collaboration. \citet{Li-influence-2018} consider  a minimum set of people influencing maximum number of people;  \citet{Demirovic-recoverable-2018} study a team that can withstand damage caused by a member leaving the team; team member replacement problem is studied by  \citet{Li-replacing-2015}.
	Majority of the papers on TFP in the  literature consider finding  one team for a given task \cite{Lappas-finding-2009,Wi-knowledge-2009,Kargar-discovering-2011,Kargar-efficient-2012,Majumder-capacited-2012,Kargar-affordable-2013,Demirovic-recoverable-2018} . Another branch of TFP deals with building teams  for multiple tasks called Multiple team foramtion problem  \cite{Gutierrez-multiple-2016,Singh-multiple-2018}. In this work, we aim to find many teams for the same task which has not been done in the literature. 

	Here  we give a comaparison of time complexities incurred by a few of the popular algorithms, namely  {\it Rarestfirst} of \citet{Lappas-finding-2009}, {\it Best leader distance}(BLD) and {\it Best sum distance} (BSD) of \citet{Kargar-discovering-2011}. 
	BLD and BSD algorithms are renamed as MinLD and MinSD by \citet{Wang-comparative-2015} which we follow in this paper. The algorithm of \citet{Kargar-discovering-2011} {\it MinLD}  takes longer than the  other algorithms since in many cases since it is constrained to consider all the nodes of the network. 
	The algorithm {\it MinSD} of \citet{Kargar-discovering-2011} considers large number of experts for each skill and that too not the rarest skill which results in  decrease of its performance. 

	The time complexities of the different algorithms that we use for comparison in this paper are tabulated in Table~\ref{time-comp-algos}.

	\begin{table}[H]
	\centering
	\begin{tabular}{|p{3.5cm}|p{3.5cm}|}
		\hline
		Algorithm&Worst case \newline Time Complexity\\ \hline
		Rarestfirst \cite{Lappas-finding-2009}& $O(|V|^2)$\\ \hline 
		CoverSteiner \cite{Lappas-finding-2009}& $O(|V|^3)$  \\ \hline
		EnhancedSteiner \cite{Lappas-finding-2009}& $O(|T||V|^2)$ \\ \hline
		MinSD \cite{Kargar-discovering-2011} & $ O(|T|^{2}V^{2}) $\\ \hline
		MinLD \cite{Kargar-discovering-2011} & $ O(|T|\times|C_{max}|\times|V|) $\\ \hline
	\end{tabular}
	\caption{Time complexity of different TF algorithms}
	\label{time-comp-algos}
	\end{table}

	\section{Motivation}\label{sec:motivation}
	Social networks are increasing in size and hence the team formation  algorithms that process the networks have to be scalable. In the case of DBLP which is a repository of publications, even if we  consider only the conferences, it can be seen that the  number of conferences increased from thousands to nearly two lakhs in the last two and half decades\footnote{https://dblp.uni-trier.de/statistics/recordsindblp}. Thus an approach that adopts a kind of divide and conquer strategy may be more profitable.

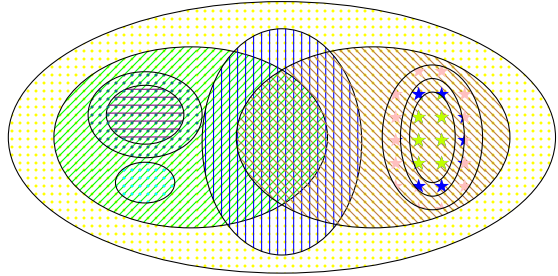
\begin{figure}[!h]
	\centering
	\begin{tikzpicture}[scale=.6]
	\draw [pattern=dots, pattern color=yellow](1,1) ellipse (6cm and 3cm);
	\draw [pattern=north east lines, pattern color=green](-1,1) ellipse (3cm and 2cm);
	\draw [pattern=vertical lines, pattern color=blue](1,.9) ellipse (1.75cm and 2.5cm);
	\draw [pattern=north west lines, pattern color=brown](3,1) ellipse (3cm and 2cm);
	\draw [pattern=dots, pattern color=cyan](-2,0) ellipse (.65cm and .45cm);
	\draw [pattern=horizontal lines, pattern color=violet](-2,1.5) ellipse (.85cm and .65cm);
	\draw [pattern=dots, pattern color=teal](-2,1.5) ellipse (1.25cm and .95cm);
	\draw [pattern=fivepointed stars, pattern color=pink](4.30,1) ellipse (1.1cm and 1.6cm);
	\draw [pattern=fivepointed stars, pattern color=blue](4.30,1) ellipse (.7cm and 1.3cm);
	\draw [pattern=fivepointed stars, pattern color=lime](4.30,1) ellipse (.5cm and 1cm);
	\end{tikzpicture}
	\caption{Modular property: Social network and its communities.}
	\label{fig:sn}
\end{figure}

	A social network exhibits modular property \cite{Newman-modularity-2006} and the ensuing clusters of nodes are called  communities. A typical social network with recursive community structure is pictorially depicted in Figure~\ref{fig:sn}. As can be seen in the picture, these communities may be overlapping, disjoint or nested communities. If teams can be extracted from each of the communities, then we get multiple teams for a task.

	\subsection{Idea of the algorithm}
	In general, while a team is being formed for, say, a hackathon, in the real world challenges, a team leader is identified at the initial step and  the team leader sets about choosing the team members who possess the necessary skills required for the task $T$ from within \textquoteleft her community', that is, with whom she has good communication. Hence as a first step in forming a team,  the team leader searches in her neighborhood. If the task is not still covered, then she looks beyond her neighbourhood to add  members for the remaining skills.

	A natural choice for a team leader is one possessing at least a few of the required skills and  having many friends.  The degree distribution of the network exhibits a typical heavy tail since the real world social networks satisfy power law. We propose that the team leader be selected from the high degree nodes in the heavy tail in order to choose highly connected person,	thus reducing the expensive computations. 


	\section{Proposed algorithm}\label{sec:algo}

	It is well known that the social networks exhibit modular property i.e. social network comprises of communities. Taking benefit of this structure, we first  identify communities, called {\it desirable communities}, containing members possessing skills relevant to the task $T$.	All skills are treated as equally important. The algorithm is given in Algorithm~\ref{algo:tfc-r} which takes as input the desirable communities computed using Algorithm~\ref{algo:dc}. If the threshold is set as 0.9,  then a desirable community must possess at least  ninety percent of skills of $T$.

		\begin{algorithm}[H]
		\caption{Algorithms DC to find desired communities}
		\label{algo:dc}
		\begin{algorithmic}[1]
			\FOR{C $ \in $ communities}
			\STATE $ cs \leftarrow \{s(v)\forall v\in C\}$ 
			\IF{$|cs\cap T| \ge threshold$}
			\STATE $ X \leftarrow $TFC-R(G(C), T)
			\ENDIF
			\ENDFOR
		\end{algorithmic}
	\end{algorithm}
	
	We propose novel algorithms, namely, TFC-R and TFC-N,  based on team leader selection who then builds the team ensuring skill coverage as well as compatibility among the team members.   TFC Algorithm~\ref{algo:tfc-r} first locates a leader within a desirable community. A leader is  defined as an expert whose degree is greater than twice the average degree of the network. Then the  other team members are searched within two hop-neighbourhood of the leader in the network so that experts covering majority of the skills of $T$ may be added. In fact, this idea is reinforced by \cite{Majumder-capacited-2012} who empirically establish that searching within two hops of a leader achieves skill coverage to a great extent and  with low communication cost. By this step, if skills of $T$ are not yet covered fully, then we add the experts in two ways.

	\subsection{TFC-R and TFC-N algorithms}
	
	In TFC-R, random experts possessing the remaining skills are added to the team. In TFC-N, nearest expert to the leader from $k$-hop neighborhood, $k>2$ is choosen.  Since the two algorithms are different only in this step, we present only the TFC-R algorithm here.

	Since social networks are very large in size, this kind of approach provides a scalable alternative to the existing algorithms which may be forced to search through the  entire network. 

\begin{algorithm}[H]
	\begin{algorithmic}[1]
	\STATE $team \leftarrow \emptyset$	
	\STATE best\_team $ \leftarrow \emptyset$
	\STATE ldbt $ \leftarrow \infty$ 
	\STATE $HD \leftarrow \{i|i \in G, ~d(i)>2\times d_{avg}, ~|s(i)\cap T|>0\}$
	\WHILE{$|HD| > 0$}
		\STATE $v \leftarrow dequeue~HD$
		\STATE $T_{NYC} \leftarrow T$
		\STATE $hop \leftarrow 1$
		\WHILE{$hop \le 2~{\bf and}~|T_{NYC}| > 0$}
			\STATE $T_{NYC} \leftarrow T$
			\STATE $team \leftarrow \emptyset$
			\STATE $team \leftarrow  team \cup \{v\}$
			\STATE $T_{C} \leftarrow s(N_{hop}(v))\cap T$
			\STATE $Nbd \leftarrow N_{hop}(v)$
			\WHILE{$|T_{C}|>0$}
				\STATE $e \leftarrow argmax_{i\in Nbd}(T_{C}\cap s(i))$
				\STATE $team \leftarrow  team \cup \{e\}$
				\STATE 	$T_{C}\leftarrow T_{C}\setminus \{s(e)\}$ 
				\STATE $T_{NYC}\leftarrow T_{NYC}\setminus \{s(e)\}$
				\STATE $Nbd \leftarrow Nbd\setminus \{e\}$
			\ENDWHILE
			\STATE $hop++ $
		\ENDWHILE
	\FOR{$skill \in T_{NYC}$}
		\STATE $re \leftarrow rand(v(skill))$ 
		\STATE $team \leftarrow  team \cup \{re\}$
	\ENDFOR
	\IF{$LD(team) < $ ldbt}
		\STATE ldbt $ \leftarrow LD(team) $
		\STATE best\_team $ \leftarrow team$
	\ENDIF
	\ENDWHILE
	\RETURN best\_team
	\end{algorithmic}		
\caption{Algorithm TFC-R}
\label{algo:tfc-r}
\end{algorithm}

	\subsection{Tracing on a toy example}
\begin{figure}[!h]
	\centering
	\begin{tikzpicture}[scale=.75]
	\node at(2,6) (A)[red,circle,draw]{A};
	\node at(2,6) (A)[green,circle,draw]{A};
	\node at(2,6.5) {A\{a\}};
	\node at (0,4)(B)[green,circle,draw]{B};
	\node [left of=B] {B\{d,e\}};
	\node at (4,4)(C)[red,circle,draw,loosely dashed]{C};
	\node at (4,4)(C)[green,circle,draw,loosely dotted]{C};
	\node [above of=C] {C\{b,c\}};	
	\node at (6,4)(D)[thick,circle,draw]{D};
	\node [right of=D] {D\{c,e\}};
	\node at (6,0)(E)[thick,circle,draw]{E};
	\node at (6,-0.5) {E\{b,c,d\}};	
	\node at (4,0)(F)[thick,circle,blue,draw]{F};
	\node [below of=F] {F\{c\}};
	\node at (0,0)(G)[thick,circle,blue,draw]{G};
	\node [below of=G] {G\{d\}};
	\node at (-2,-2)(H)[circle,draw]{H};
	\node at (-2,-2.5) {H\{c,e\}};
	\node at (-2,0)(I)[thick,circle,blue,draw]{I};
	\node at (-2,0.5) {I\{e\}};
	\node at (2,-1)(J)[thick,blue,circle,draw]{J};
	\node at (2,-1.5) {J\{a,b\}};
	\node at (0,-2)(K)[circle,draw]{K};
	\node at (0,-2.5) {K\{a\}};
	\node at (4,-2)(L)[circle,draw]{L};
	\node at (4,-2.5) {L\{b\}};
	\node at (7,2)(M)[circle,draw]{M};
	\node at (7,2.5) {M\{b\}};
	\node at (8,0)(N)[circle,draw]{N};
	\node at (8,-0.5) {N\{c\}};
	\node at (5,2)(O)[circle,draw]{O};
	\node at (5,2.5) {O\{d\}};
	\node at (6,6)(P)[circle,draw]{P};
	\node at (6,6.5) {P\{d\}};
	\node at (4,7)(Q)[green,circle,draw]{Q};
	\node at (4,7.5) {Q\{b,e\}};
	\node at (0,7)(R)[circle,draw]{R};
	\node at (0,7.5){R\{a\}};
	\node at (2,4)(S)[red,circle,draw]{S};
	\node at (2,3.5) {S\{d,e\}};
	\node at (2,2)(T)[circle,draw]{T};
	\node at (2,1.5) {T\{d\}};
	
	\draw [solid,green,thick,dashed] (A) to node[above]{4} (B);
	\draw [solid] (B) to node[above]{2} (T);
	\draw [solid] (C) to node[above]{2} (T);
	\draw [solid,red]decorate [decoration={snake}]{(C) to node[above]{3} (S)};
	\draw [solid,green,thick,dashed] (A) to node[above]{2} (C);
	\draw [solid,red]decorate [decoration={snake}] {(A) to node[above]{2} (C)};
	\draw [solid] (B) to node[left]{3} (G);
	\draw [solid] (C) to node[right]{4} (F);
	\draw [solid] (G) to node[above]{3} (H);
	\draw [solid] (G) to node[above]{3} (F);
	\draw [solid,blue] (G) to node[above]{2} (I);
	\draw [solid] (C) to node[above]{3} (D);
	\draw [solid] (D) to node[right]{3} (E);
	\draw [solid] (E) to node[above]{2} (F);
	\draw [solid,blue] (G) to node[below]{3} (J);
	\draw [solid,blue] (F) to node[below]{2} (J);
	\draw [solid] (L) to node[below]{2} (J);
	\draw [solid] (J) to node[below]{1} (K);
	\draw [solid] (O) to node[above]{3} (E);
	\draw [solid] (N) to node[above]{2} (E);
	\draw [solid] (M) to node[above]{3} (E);
	\draw [solid] (A) to node[below]{2} (R);
	\draw [solid,green,thick,dashed] (A) to node[below]{1} (Q);
	\draw [solid] (P) to node[left]{1} (D);
	
	\end{tikzpicture}
	\caption{Toy example}
	\label{toy}
\end{figure}
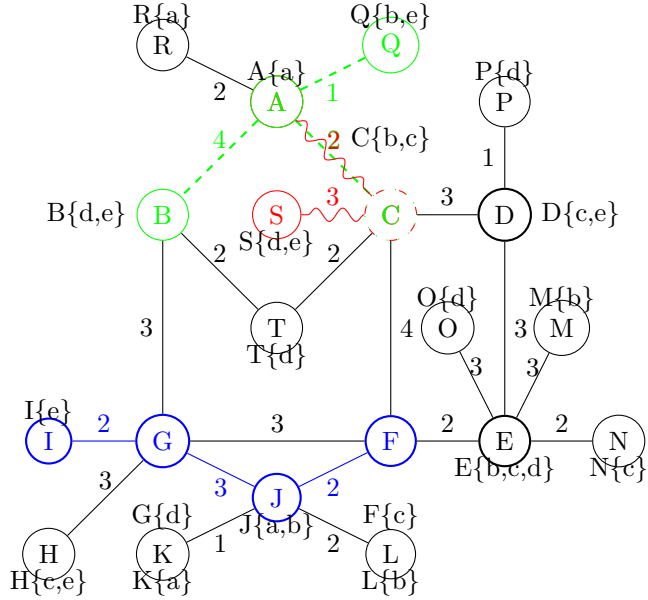

In order to compare and contrast the proposed algorithm with some of the popular ones in the literature, we design  a toy example as shown in Figure~\ref{toy}. 
Skill-centric algorithms like Rarestfirst\cite{Lappas-finding-2009} and MinLD, MinSD of \cite{Kargar-discovering-2011}, design heuristics that start with a skill and find expert who has the least communication distance to the team built so far. We can appreciate the difference in the approaches of the different algorithms by looking at the Toy example\ref{toy}.

We trace the algorithm on the  toy example in Figure~\ref{toy} and the comparative results are given in Table~\ref{toy-results}. For a task requiring skills $\{a, b, c, d, e\}$, the teams $ABCQ$ and $DEFJ$ tie in terms of cardinality. Consider node $C$, whose closest experts with skill $d$  are  $S$ and $T$. Since  $T$ is at least distance from $C$,  the algorithms Rarestfirst,  MinSD and MinLD choose $T$. But this choice increases cardinality.  TFC-R chooses  $S$ instead of $T$, thus  obtaining the team \{A, C, S\}. Hence the expert-centric algorithms  result in  smaller teams.  

\begin{table}[!h]
	\centering
	\begin{tabular}{|l|l|c|c|c|c|}
		\hline
		\multirow{2}{*}{Algorithm}&\multirow{2}{*}{Team (Leader)}&\multicolumn{3}{|c|}{distance metrics}\\ \cline{3-5}
		&&Diameter&SD&LD \\
		\hline
		TFC-R       & ACS (C)    &   2   &   26  &   5   \\     \hline
		Rarestfirst & JFGI (J)   &   2   &   30  &   10  \\     \hline
		MinSD       & ACS (S)    &   2   &   26  &   8   \\     \hline
		MinLD    	& AQCB (A)   &   3   &   28  &   7  \\		\hline
	\end{tabular}
	\caption{Cost of teams obtained by Rarestfirst, MinSD, MinLD and TFC-R algorithms on the toy example}
	\label{toy-results}
\end{table}

\section{Benchmark dataset}\label{sec:benchmark}
We curated a large collaboration network from DBLP database\footnote{https://dblp.uni-trier.de/xml/}. 	
DBLP data is modeled as a social network with experts as nodes and distance based on mutual collaboration denoting the edge weight giving rise to an undirected and weighted network. We have followed the modeling method exactly as suggested originally by Lappas et al.~\cite{Lappas-finding-2009} and followed by \cite{Aris-power-2010,Kargar-discovering-2011,Majumder-capacited-2012,Kargar-efficient-2012,Aris-online-2012}. The snapshot of DBLP data set considered has been restricted to conferences pertaining to four major research areas of  computer science: Database(DB), Data Mining(DM), Artificial Intelligence(AI) and Theory(TH). Data comprises of publications in the conferences specified in Table~\ref{DBLP-conf}. We treat the data of each research area as a bigger community and the publications in each conference  as  smaller community. Details of the total benchmark data set is given in Table~\ref{ns} of Appendix.
\begin{table}
	\centering
	\begin{tabular}{|c|l|}
		\hline
		Research area & Conferences\\ \hline
		\multirow{2}{*}{DB} & VLDB, SIGMOD, ICDT, \\
		&ICDE and PODS\\ \hline
		\multirow{2}{*}{DM} & WWW, SDM, KDD, ICDM, \\
		&PKDD and WSDM\\ \hline
		\multirow{2}{*}{AI}  & NIPS, IJCAI, ICML, UAI, \\
		&COLT and CVPR\\ \hline
		\multirow{2}{*}{TH} & FOCS, SODA, STOC, ICALP, \\
		&STACS and ESA\\ \hline
	\end{tabular}
	\caption{Research areas are taken as big communities and constituting conferences as small communities}
	\label{DBLP-conf}
\end{table}

Individuals who have published papers in the conferences are considered as authors. Authors having at least three publications  are considered as experts. 	
Now to associate skills to each expert, the words in the  titles of the  publications are utilized.  Each title is segmented into constituent words then from these words, non-trivial words are extracted by removing stop words. 
Roots of the  words are retained using the stemming procedure available in Natural Language Tool Kit(NLTK)\footnote{https://www.nltk.org}.  The words  that appear at least twice in the publication titles of an expert are considered as skills possessed by the expert. Two experts $ v_{i}$ and $ v_{j}$  are treated as collaborating nodes if only if they have minimum three joint publications indexed in DBLP. 
The amount of collaboration  between  $ v_{i}$ and $ v_{j}$  is calculated by using Jaccard distance measure which is taken as  edge weight $e_{ij}$, given by  1 $ -\frac{| P_{v_{i}} \cap  P_{v_{j}}| }{| P_{v_{i}} \cup P_{v_{j}}| } $ where $ P_{v_{i}} $ represents the number of papers published by $ v_{i} $. Hence in the network lesser the Jaccard distance is, more their collaboration and vice-versa.

The data sets considered by the authors  so far in the literature  are small in size. Number of nodes in these networks is less than ten thousand. In this work we build a network  containing more than 30,000 authors and nearly 98,000 edges.  The details of the data sets that have been  considered by the other papers as well as ours are given in Table~\ref{comp-datasets}. 

\begin{table}[!h]
	\centering
	\begin{tabular}{|l|l|l|l|l|l|}
		\hline
		Algorithm&$|V|$&$|E|$&$|S|$&$|E|/|V|$\\ \hline
		Rarestfirst\ \cite{Lappas-finding-2009} & \multirow{2}{*}{5508} & \multirow{2}{*}{5588} & \multirow{2}{*}{1792} & \multirow{2}{*}{1.014}\\ 
		Enhancedsteiner\ \cite{Lappas-finding-2009}&&&&  \\ \hline
		GenTeam\ \cite{Li-generalized-2010}&5482&10339&11905&1.885\\ \hline
		MinSD\ \cite{Kargar-discovering-2011} &\multirow{2}{*}{5658}&\multirow{2}{*}{8588}& \multirow{2}{*}{-} &\multirow{2}{*}{1.517}\\
		MinLD\ \cite{Kargar-discovering-2011}&&&&\\ \hline
		MinDiaSol\ \cite{Majumder-capacited-2012}&7159&15110&4355&2.110\\ \hline
		MinMaxSol\ \cite{Majumder-capacited-2012}&6229&9400&-&1.509 \\ \hline
		TF survey$^{*}$ \cite{Wang-comparative-2015}&7332&19248&2763&2.625\\ \hline		
		TFC-R,N & \textbf{32477}& \textbf{98676}&\textbf{13232}&\textbf{3.038}\\ \hline
	\end{tabular}
	\caption{ DBLP data set taken by TFC-R is much larger than those considered by the other algorithms in the literature. TF survey$^{*}$ is not an algorithm}
	\label{comp-datasets}
\end{table}

\subsection{Problem setting}

The problem is to find a team of authors who are capable of performing the task of writing  a given research paper. The non-trivial words from the title of the paper are considered as skills required and the problem is to find experts from the DBLP data set having the required skills and incurring minimum communication cost. 


\section{Network analysis of DBLP network}\label{sec:network-analyis}

A preliminary network analysis of the DBLP network is conducted on (i) connected components, (ii) degree distribution and (iii) the neighbourhoods of  high degree nodes from the perspective of choosing experts with the required skills for team formation. The insights gained from this analysis that  helped us design TFC-R are explained in this section.

\subsection{Connected components in DBLP }

\begin{figure}[!h]
	\centering	
	\includegraphics[scale=.65]{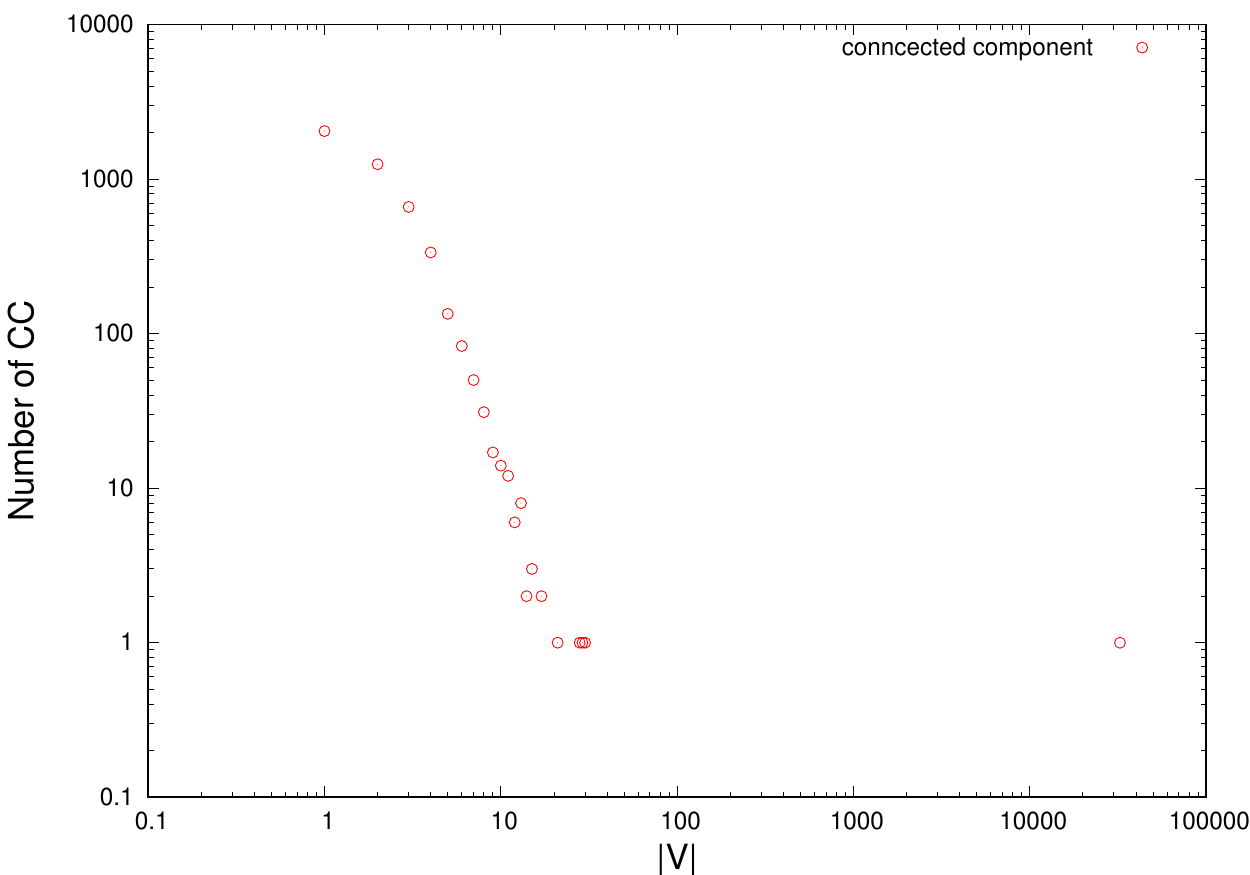}
	\caption{\textbf{Power law}: Connected components satisfying power law for DBLP network}
	\label{cc}
\end{figure}

It is interesting to note that the connected components of the DBLP network satisfy power law as seen in the Figure~\ref{cc}. That is, there exist large number of connected components having less number of nodes and smaller number of components having a large number of nodes, Greatest Connected Component (GCC) being one of them. In this context, let us analyze the existing algorithms in the literature, most of whose initial step is to choose an expert with rarest skill. By empirical analysis we see that if the initial expert falls within one of the smaller components, the search for other experts takes longer time as well as the cost incurred may be higher as the other experts lie at longer distances from the initial expert. Hence it is useful to start the search in GCC of the network which the TFC-R, TFC-N algorithms carry out.   

\subsection{Degree distribution}

\begin{figure}[!h]
	\centering	
	\includegraphics[scale=.65]{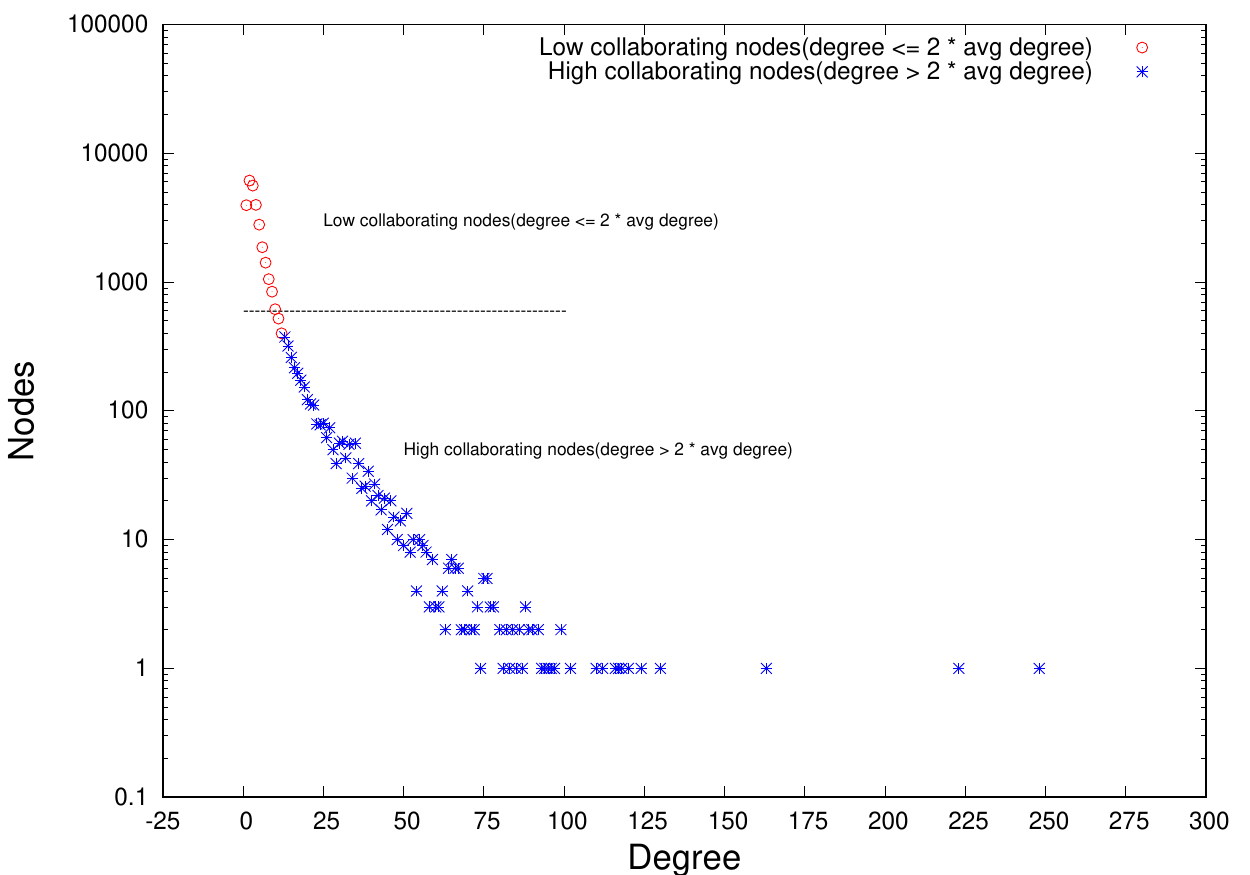}
	\caption{\textbf{Power law}: Degree distribution satisfying power law for DBLP network}
	\label{dd}
\end{figure}

DBLP network is a typical social network whose degree distribution follows power law as can be seen in Figure~\ref{dd}. Once again for the same reason, if the rarest expert is one of the low degree nodes, the search for the team becomes longer and hence it will be useful to start with a higher degree node.  It is found that the ratio of high collaborating (having degree higher than twice the average degree of the community) nodes to low collaborating nodes in DBLP network is 1:9(3293:29184). Hence choosing a high collaborating node as a leader is a good starting point for the Team formation algorithm.

\subsection{Neighbourhoods of high degree nodes}

\begin{figure}[!h]
	\centering	
	\includegraphics[scale=.65]{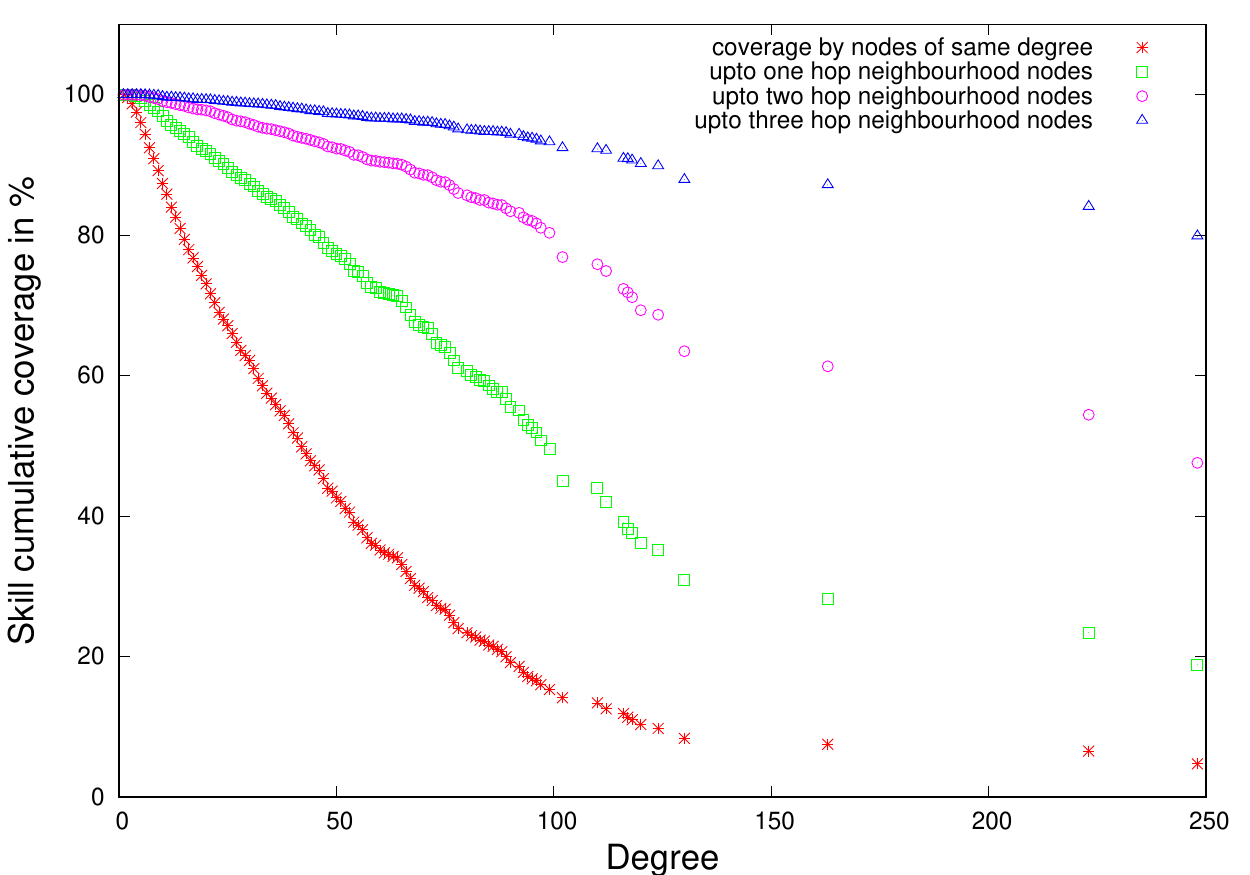}	
	\caption{Percentage of skills covered cumulatively by the node alone and by adding 1-hop, 2-hop and 3-hop neighbourhoods.}
	\label{cskc}
\end{figure}

Let skill coverage of an expert node be the skills possessed by the expert.
The plot given in Figure~\ref{cskc} shows an interesting structure of the social network from the point of view of skill coverage.  For each node of degree $k$, though its skill coverage may be low, by including its 1-hop and 2-hop neighbours improves the skill coverage between 90 to 100\%. 

The proposed Team Formation Algorithm TFC-R starts by choosing an expert from high collaborating nodes, one who possesses at least one skill required for the task, as team leader,  then adds the other  team members from unit hop neighbourhood and by successively increasing the hop length up to 2 or 3 steps. 
It is evident from Figure~\ref{cskc}, that the two hop neighborhood of high degree nodes having degree greater than average degree covers above 90\% of community skills.  Skill coverage by 3-hop neighborhood alone does not add too much. Therefore, we can limit the neighbourhood  up to 2-hops i.e. (immediate friend network and friend's friend network) for team formation.


The  	results obtained by the proposed algorithms are discussed in detail in the next section. 

\section{Implementation and results}\label{sec:results}
The algorithms have been  implemented in Python on Intel(R) Core(TM) i5-4300M CPU @ 2.60GHz. 
In this paper, it is important to note that,  we do not retrieve communities using community discovery algorithms, but consider the subsets of DBLP based on research areas like DB, DM, VLDB etc as communities. 	
The largest connected components(LCC) of DBLP, DB and VLDB networks have been considered for experimentation. 

We carry out 100 experiments for each $k$ by randomly choosing subsets of size $k$ as tasks $T$, for $k = 4 ,5,\ldots 20$ and the algorithms are implemented on these data sets to find  teams for T. The average cost/cardinality of the teams obtained for the 100 tasks is tabulated for each $k$. We compare TFC-R and TFC-N algorithms with respect to the communication cost measures of {\it diameter}, {\it sum distance} and {\it leader distance}  against MinSD,  MinLD$^{*}$ and RF$^{*}$ algorithms. The asterik indicates modified faster implementations. Since MinLD considers every node as a leader and hence  takes too long a time, it has been modified to consider only those nodes having degree greater than twice the average degree as leaders. The modified implementation by \citet{Wang-comparative-2015} is taken for Rarestfirst algorithm.

Further, in order to keep the comparison fair, all the algorithms have been implemented on VLDB network. We also compare the performance of TFC-R algorithm on the  entire DBLP  network in relation to that of its performance on the communities of DBLP network (DB and VLDB). 	The datasets and code are available in public domain\footnote{https://github.com/abrameshba/teamformation}

The results are organized as follows:
\begin{enumerate}
	\item[(A)] Execution time of the algorithms
	\item[(B)] Comparison of Team size obtained by the algorithms 
	\item[(C)] Comparison of the algorithms with respect to the communication costs 
	\begin{enumerate}
		\item[I] diameter distance
		\item[II] leader distance
		\item[III] sum distance
	\end{enumerate}
	\item[(D)] Scaling of the results as the experiments are carried over each of the sets DBLP, DB and VLDB. Note that VLDB is a subset of DB which is a part of the whole network DBLP.
\end{enumerate}

\subsection{Execution time}	
\begin{figure}[!h]
	\centering
	\includegraphics[scale=.65]{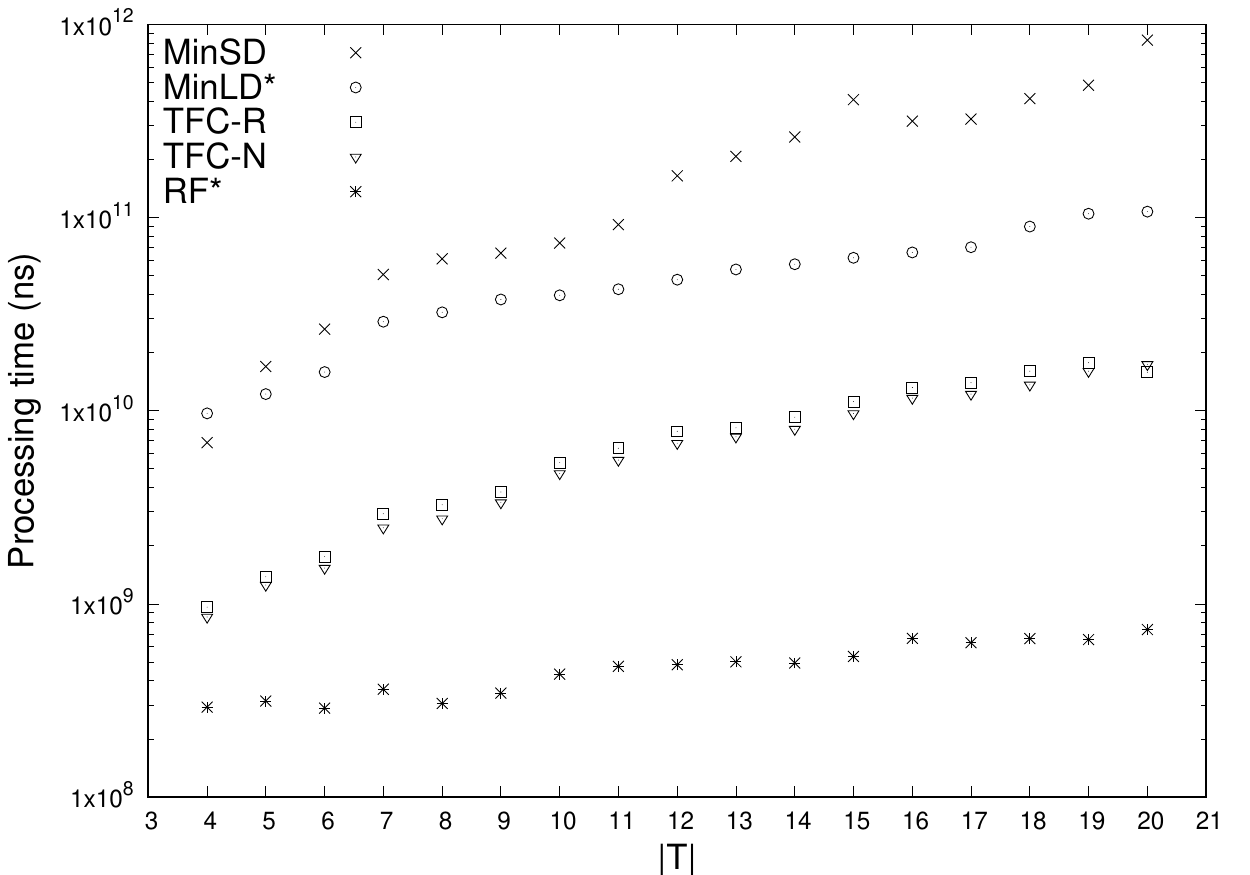}
	\caption{Comparison of average processing time taken by TFC-R, TFC-N,  Rarestfirst(RF), MinLD and MinSD algorithms for VLDB network. *Note that  RF$^{*}$ and MinLD$^{*}$ are modified and faster implementations.   }
	\label{all-processing}
\end{figure}

The faster version of Rarestfirst  RF$^{*}$ gives the fastest results, followed closely by TFC-R and TFC-N. Clearly both the proposed algorithms are several order faster than both MinSD and MinLD$^{*}$ as seen in Figure~\ref{all-processing}.

\subsection{ Team size }

\begin{table}
	\centering
	\begin{tabular}{|l|l|l|l|l|l|}\hline
		$|T|$&RF$ ^{*} $&MinLD$ ^{*} $&MinSD&TFC-R&TFC-N \\ \hline
		4&3.87&\textbf{3.27}&3.87&3.82&3.81 \\ \hline
		5&4.84&\textbf{4.19}&4.77&4.76&4.75 \\ \hline
		6&5.82&\textbf{5.19}&5.74&5.67&5.65 \\ \hline
		7&6.73&\textbf{6.08}&6.62&6.57&6.56 \\ \hline
		8&7.73&\textbf{6.92}&7.5&7.34&7.3 \\ \hline
		9&8.69&\textbf{8.01}&8.52&8.3&8.28 \\ \hline
		10&9.55&\textbf{8.8}&9.27&8.99&9.02 \\ \hline
		11&10.4&\textbf{9.68}&10.24&9.69&9.69 \\ \hline
		12&11.37&\textbf{10.44}&11.09&10.47&10.47 \\ \hline
		13&12.29&11.39&11.98&11.4&\textbf{11.33} \\ \hline
		14&13.12&12.38&12.9&12.28&\textbf{12.18} \\ \hline
		15&14.07&13.11&13.65&13.07&\textbf{12.97} \\ \hline
		16&14.85&13.89&14.45&13.77&\textbf{13.75 }\\ \hline
		17&15.66&14.75&15.36&14.64&\textbf{14.49} \\ \hline
		18&16.47&15.51&16.08&15.11&\textbf{15.05} \\ \hline
		19&17.59&16.43&17.16&16.1&\textbf{15.99 }\\ \hline
		20&18.53&17.38&18.01&16.76&\textbf{16.73} \\ \hline
	\end{tabular}
	\caption{Average cardinality of teams given by TFC-R, TFC-N, RF$^{*}$, MinSD and MinLD$^{*}$ algorithms on VLDB network}
	\label{DB_cardinality}
\end{table}

As seen in Table~\ref{DB_cardinality}, the average sizes obtained by TFC-N are better for tasks of size greater than 12 and MinLD for smaller tasks. Both TFC-R and TFC-N give far better team sizes  when compared to RF$^{*}$ as well as MinSD.

\subsection{Communication cost: diameter}	
\begin{table}
	\centering
	\begin{tabular}{|l|l|l|l|l|l|}\hline
		$|T|$&RF$ ^{*} $&MinLD$ ^{*} $&MinSD&TFC-R&TFC-N \\ \hline
		4&5.99&\textbf{5.25}&5.75&6.2&5.63 \\ \hline
		5&6.9&\textbf{6.1}&6.82&7.3&6.58 \\ \hline
		6&7.71&\textbf{6.78}&7.43&8.05&7.18 \\ \hline
		7&8.08&\textbf{7.12}&7.83&8.49&7.43 \\ \hline
		8&7.96&\textbf{7.11}&7.83&8.58&7.39 \\ \hline
		9&8.95&\textbf{7.91}&8.58&9.22&8.2 \\ \hline
		10&8.57&\textbf{7.7}&8.15&8.82&7.77 \\ \hline
		11&8.86&\textbf{7.8}&8.35&9.08&7.95 \\ \hline
		12&9.37&\textbf{8.13}&8.63&9.37&8.29 \\ \hline
		13&9.77&\textbf{8.66}&9.01&9.97&8.76 \\ \hline
		14&9.57&\textbf{8.6}&8.97&10.09&8.81 \\ \hline
		15&9.71&\textbf{8.78}&9.31&10&8.8 \\ \hline
		16&9.97&\textbf{9.17}&9.48&10.5&9.31 \\ \hline
		17&9.84&\textbf{8.96}&9.22&10.36&9.03 \\ \hline
		18&10.19&\textbf{8.99}&9.22&10.45&9.07 \\ \hline
		19&10.14&9.07&9.25&10.59&\textbf{9.06} \\ \hline
		20&10.26&\textbf{9.41}&9.86&10.98&9.59 \\ \hline
	\end{tabular}
	\caption{Average diameter distance of teams generated by TFC-R, TFC-N, RF$^{*}$, MinSD and MinLD$^{*}$ algorithms on VLDB network}
	\label{all_diameter}
\end{table}

Table~\ref{all_diameter} shows that the best performance is obtained by MinLD$^{*}$ closely followed by TFC-N. Our proposed algorithm TFC-N gives teams with better average diameter when compared to  all the other algorithms of RF$^{*}$, MinSD as well as TFC-R.  The diameter of the teams obtained by   TFC-R algorithm seem slightly inferior in comparison as TFC-R may be  adding a distant random expert. 

\subsection{Communication cost: leader distance}		
\begin{figure}[!h]
	\centering
	\includegraphics[scale=.65]{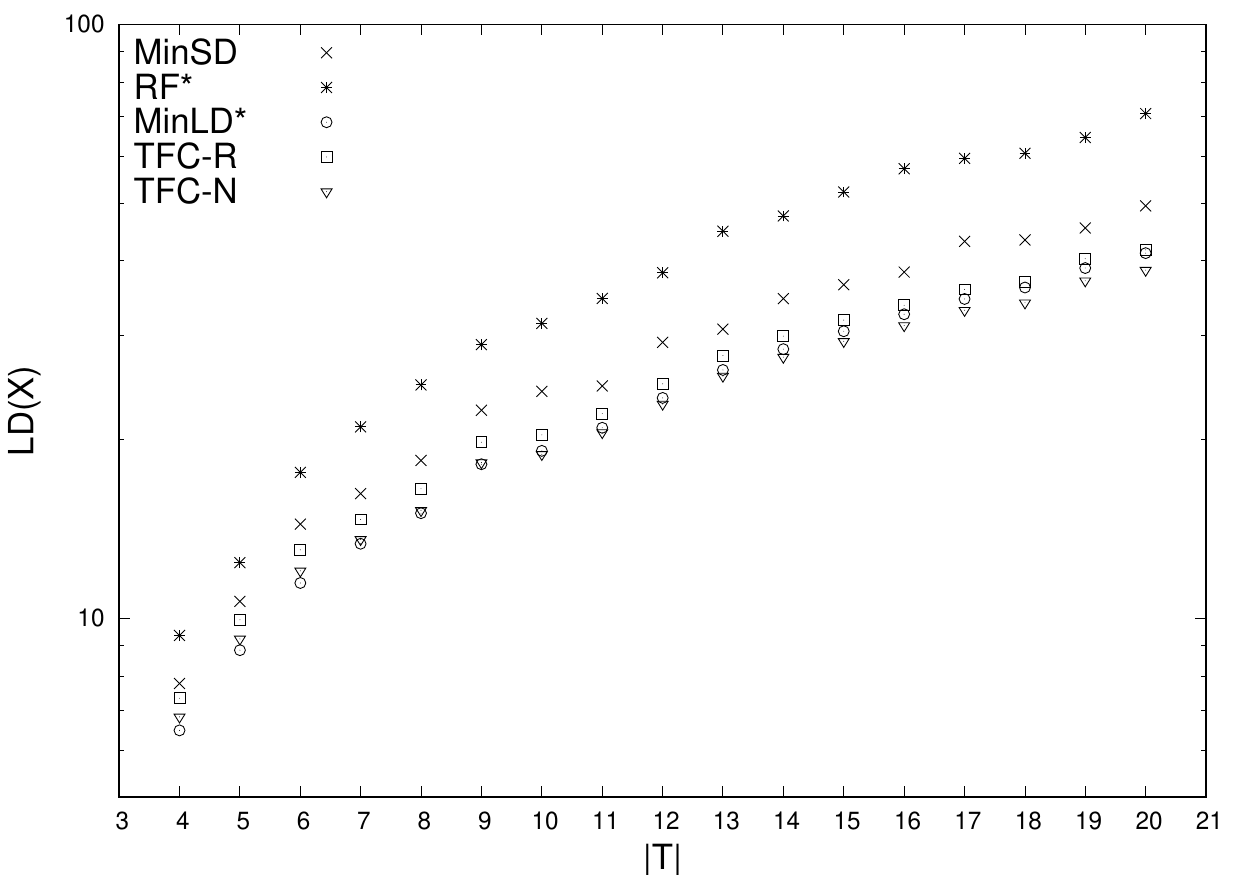}
	\caption{Comparison of  average \textit{leader distance} of teams obtained by TFC-R, TFC-N, Rarestfirst*, MinSD, MinLD* algorithms on VLDB network}
	\label{all_leader_distance}
\end{figure} 

The results in Figure~\ref{all_leader_distance} show that MinLD, TFC-R and TFC-N are performing well with respect to leader distance measure.
Rarestfirst  and MinSD algorithms are not based on choosing a  leader hence do not perform well and Rarestfirst algorithm  does not seem to scale well as the size of the task increases. 
\subsection{Communication cost: sum distance}		
\begin{figure}[!h]
	\centering
	\includegraphics[scale=.65]{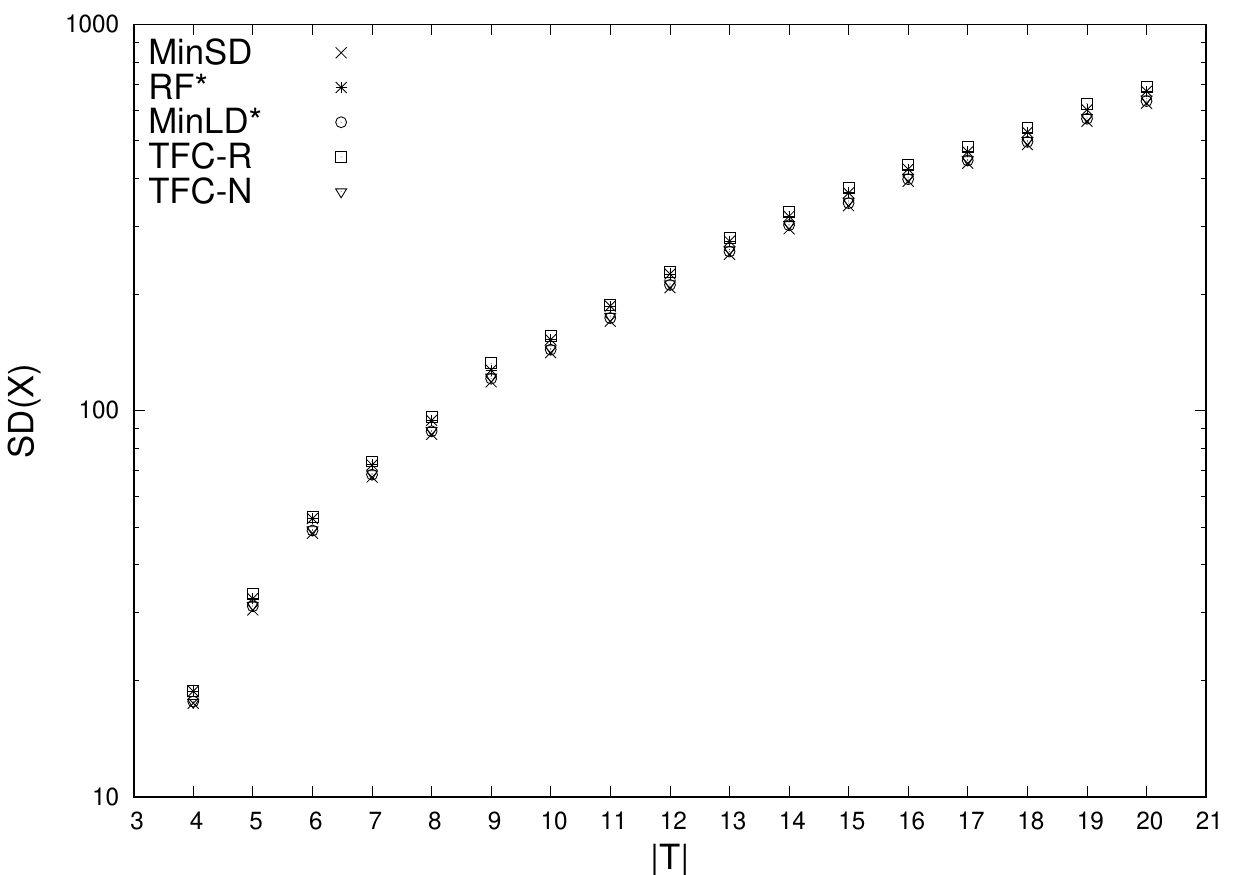}
	\caption{The performance of  TFC-R, TFC-N, Rarestfirst*, MinSD, MinLD* algorithms w.r.t. average \textit{sum distance} in VLDB network}
	\label{all_sum_distance}
\end{figure}

The plot in Figure~\ref{all_sum_distance}  indicates that all the algorithms have similar performance with respect to the sum distance measure.

\section{Performance of TFC-R on communities}\label{sec:results-comm}

In this section, the algorithms TFC-N and TFC-R are studied closely for their scalability and correctness with respect to the search being restricted to a community. The results are presented for TFC-R and these are the most attractive experiments that show-case the importance of the algorithm with respect to its scalability for large networks.

All the tasks are selected from VLDB community. For each task skills are randomly selected. Note that the  VLDB network is contained in DB and DB in DBLP. Each experiment is repeated 100 times for all the tasks of sizes varying between 4 to 20 and the average results are tabulated.  The experiments for tasks of size greater than 16 have not been conducted on  DBLP since 100 experiments have to conducted where each experiment takes more than 30 minutes for each task.

\subsection{Processing time}
\begin{figure}[!h]
	\centering
	\includegraphics[scale=.65]{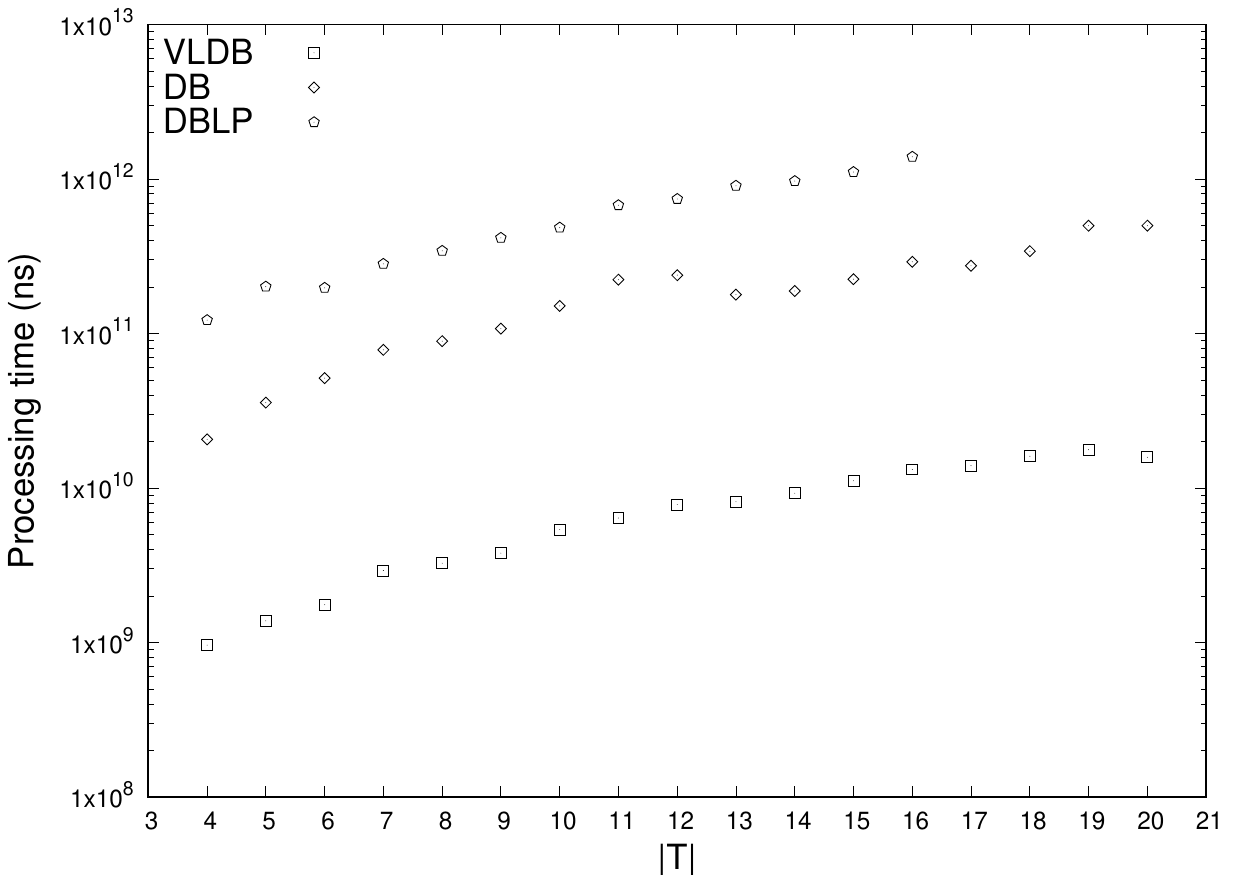}
	\caption{Average processing time taken by TFC-R for DBLP, DB and VLDB networks}
	\label{tfc-processing}
\end{figure}

Clearly this is the most important experiment as it shows that the algorithm is much faster when run on smaller communities by orders of magnitude in comparison to the larger network. The plot in Figure~\ref{tfc-processing} shows that the algorithm runs between 50 to 100 times faster on average on VLDB when compared to DB and DB is clearly 10 times faster than when run on the whole network.


\subsection{Cardinality}
\begin{figure}[!h]
	\centering
	\includegraphics[scale=.65]{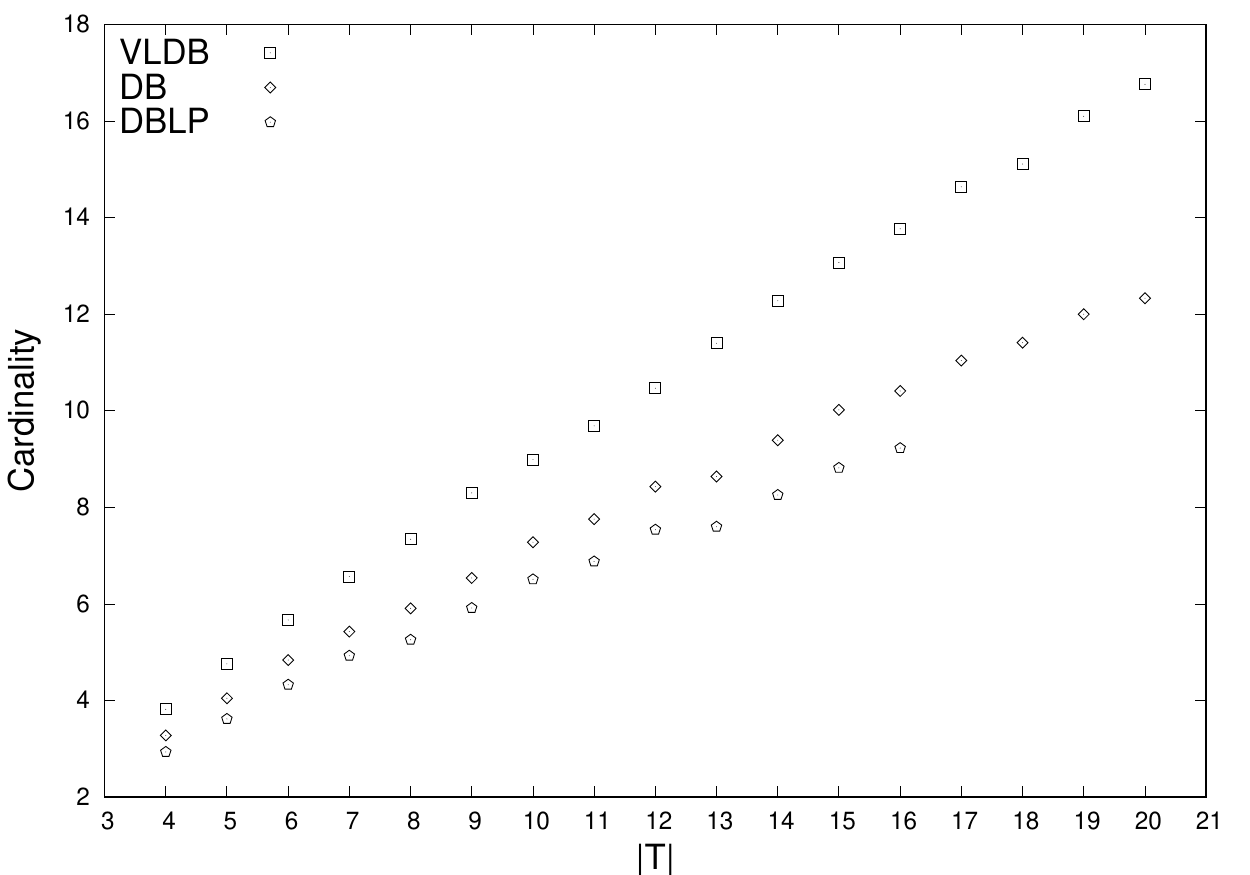}
	\caption{Cardinality comparison of TFC-R on DBLP, DB, VLDB}
	\label{tfc_cardinality}
\end{figure}

In this experiment, we would like to test the size of the team obtained by the TFC algorithm when tested on different communities: VLDB which is smaller than DB  and on the whole network DBLP. The team found if the search is restricted to the smaller VLDB network naturally may be of larger size, in comparison to when the search is expanded to DB and then DBLP. As seen in Figure~\ref{tfc_cardinality}, the cardinality of teams found in smaller communities  are slightly bigger than those found in the whole of DBLP, but the difference is negligible. Hence this algorithm would be very useful in the case of large networks since it can find teams from within communities of the network.

\subsection{Random experts}

\begin{figure}[!h]
	\centering
	\includegraphics[scale=.65]{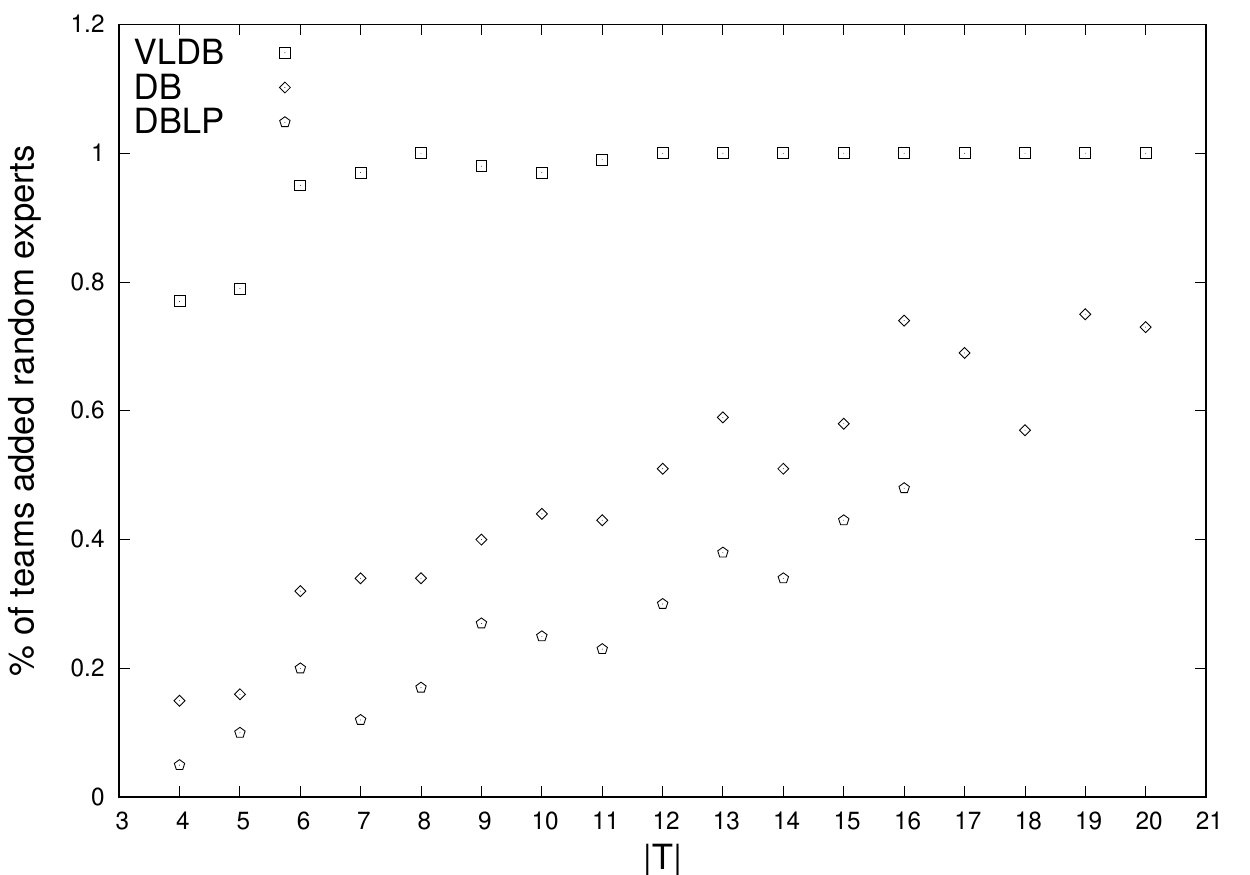}
	\caption{The  (\%) of teams in which random experts are  required to be added for TFC on DBLP, DB and VLDB}
	\label{tfc_rand}
\end{figure}

TFC-R is almost like the reverse of Rarestfirst algorithm. In Rarestfirst algorithm  expert possessing rarest skill is first added to the team whereas TFC adds rarest skilled person last. A random expert possessing rarest skill is added if he/she explicitly is required to be added to the team. Figure~ \ref{tfc_rand} indicates that the number of teams having a  random expert is high for the small network of VLDB. On the other hand, only 60- 70\% of the teams on average require random expert to be added for the larger networks of DB and DBLP. 


\subsection{Communication cost}

\begin{figure}[!h]
	\centering
	\includegraphics[scale=.65]{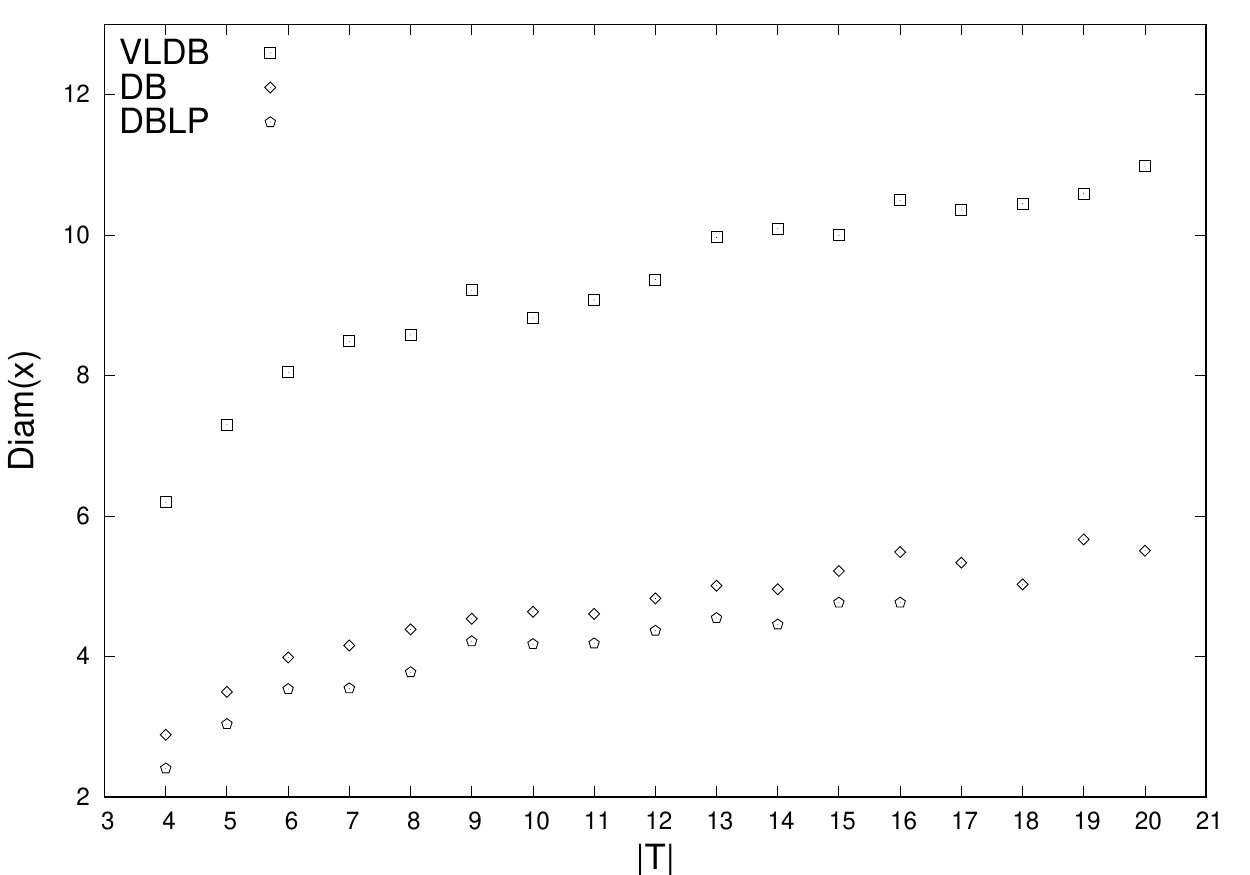}
	\caption{Comparison of  TFC-R on DBLP, DB and VLDB with respect to \textit{diameter distance} measure}
	\label{tfc_diameter}
\end{figure}

\begin{figure}[!h]
	\centering
	\includegraphics[scale=.65]{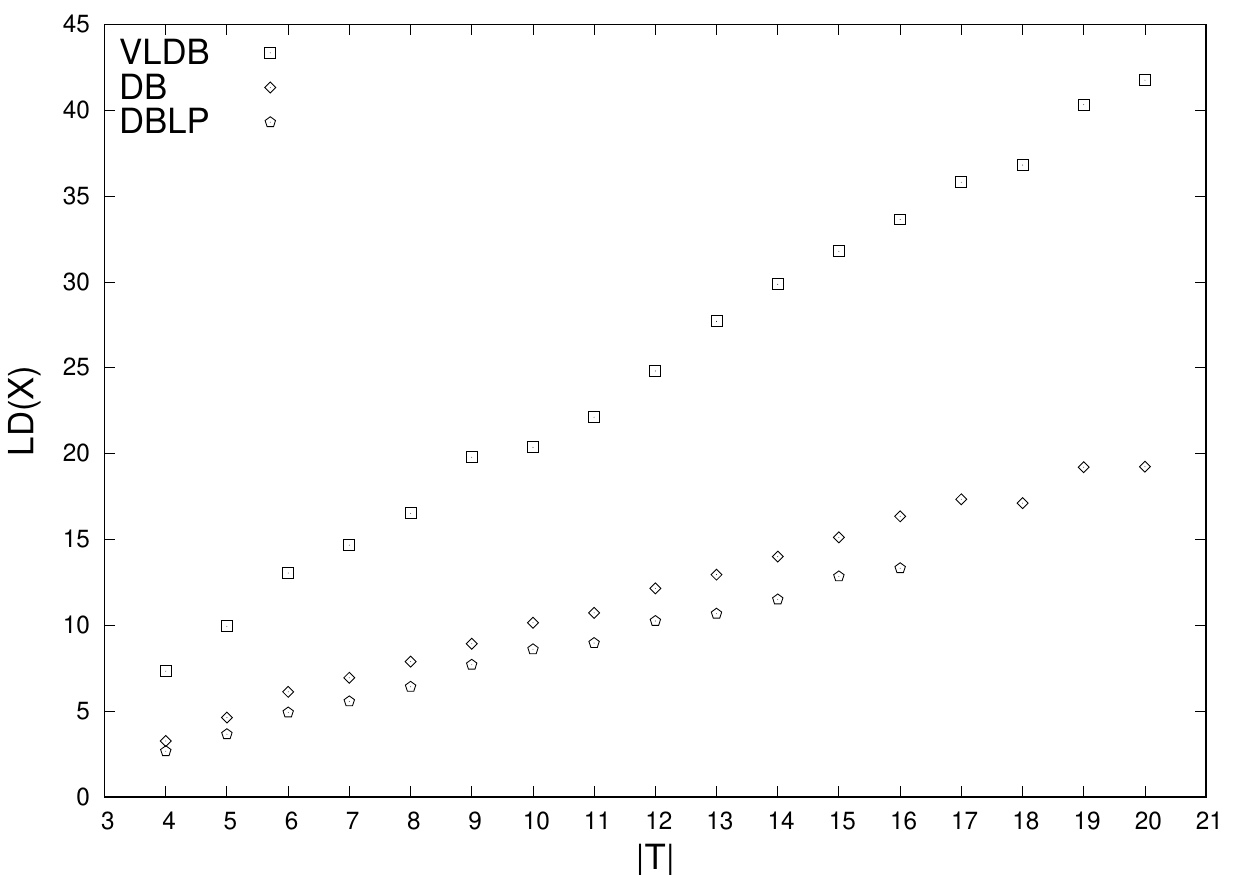}
	\caption{Comparison of TFC-R on DBLP, DB and VLDB for average \textit{leader distance}}
	\label{tfc_ld}
\end{figure}

\begin{figure}[!h]
	\centering
	\includegraphics[scale=.65]{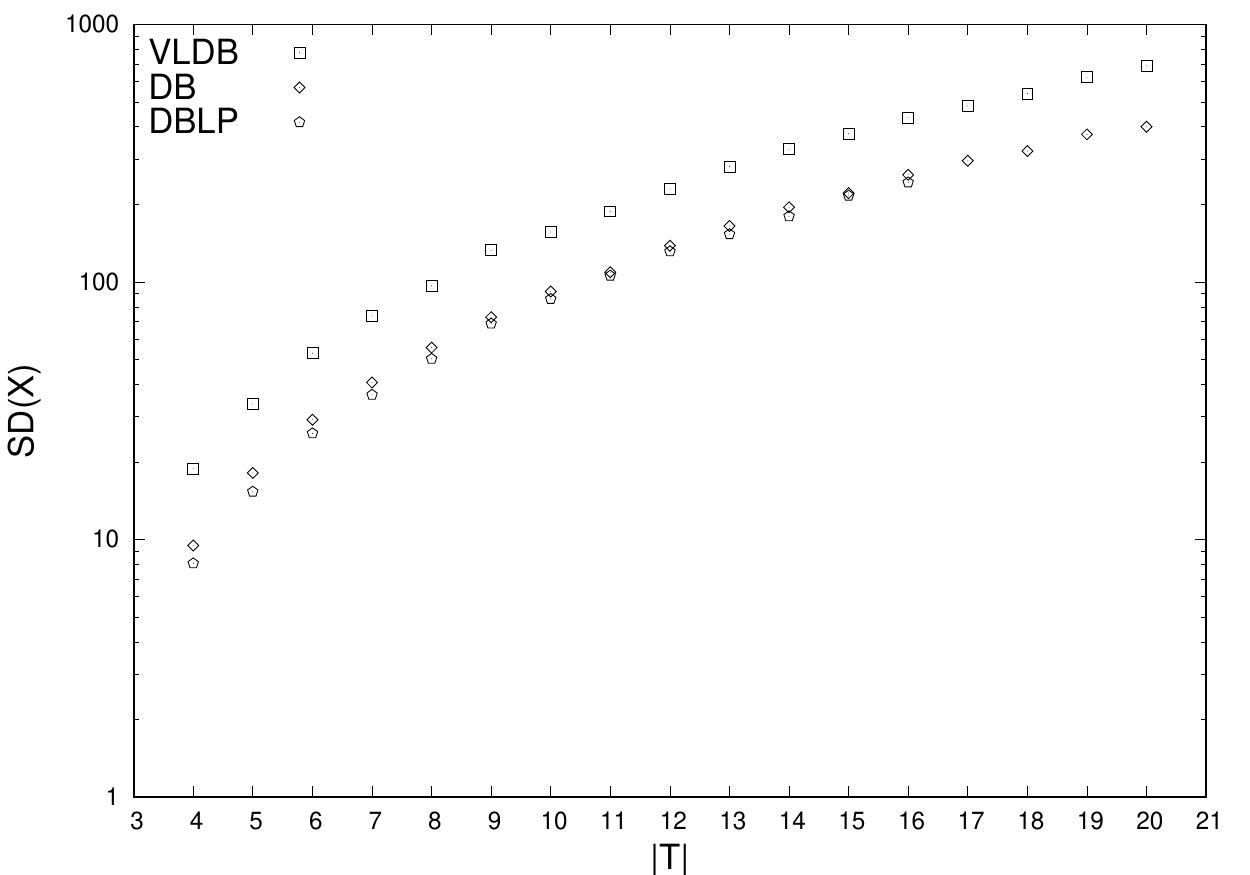}
	\caption{Comparison of average \textit{sum distance} values of teams obtained by TFC-R on DBLP, DB and VLDB}
	\label{tfc_sd}
\end{figure}

The performance of the TFC algorithm is assessed by searching for teams within different communities: small one like VLDB, a bigger one like DB and the whole network of DBLP. The performance is evaluated by computing the communication costs of diameter, leader distance and sum distance. As can be seen in Figures ~\ref{tfc_diameter}, \ref{tfc_ld} and \ref{tfc_sd}, the costs are  slightly higher for teams retrieved from the small community of VLDB in comparison to those from the bigger communities of DB and the whole network DBLP. On the other hand, the teams found in DB have almost the same cost as those of the whole network.

\section{Case study}\label{case-study}

In this section, we show-case the ability of TFC-R (and TFC-N) algorithm for finding multiple teams for a task with less communcation cost. 
We considered three papers that have achieved  best paper awards in  VLDB  given in \cite{Huang-2020} that is chosen as part of the case-study by \cite{Wang-comparative-2015}. These translate to three tasks $T_1, T_2$ and $T_3$ with 5,8 and 10 skills respectively. The actual authors of these papers  are listed in Table \ref{case_tasks}. 

Table \ref{case_results} shows the teams identified community-wise by TFC-R algorithm.
The communities given here are nested, overlapping or disjoint as shown in Figure~\ref{fig:sn}. The research paper belongs to a particular conference contained in a community, it will be interesting to find teams when the search is carried out in the other desirable communities as well as the entire social network. 	
Clearly the experts who have high collaborations within the communities like Samuel Madden have been identified by TFC-R for task 2 in all the three networks of VLDB, DB and DBLP. Madden also happens to be one of the authors of the paper given for task 2.
The communities which do not have the necessary skill coverage are kept blank.
It can be seen that experts with high collaboration and having research in multiple domains have been identified for tasks 1 and 3. Experts marked with $^{*}$ are authors who are added in the random step who may not have come up in the 2-hop neighbourhoods in the first instance.



This case-study shows that for some tasks the high degree nodes make the teams into singleton sets (single-author papers). Since all the tasks are from database research area, we find that interesting teams can be found in non-database communities like AI, Theory(TH), STOC etc.  In fact, it may be  more interesting, if we exclude the high degree author, in order to obtain teams with diversity.


\begin{table}[h]
	\centering
	\begin{tabular}{|p{.05\linewidth}|p{.35\linewidth}|p{.35\linewidth}|p{.25\linewidth}|}
		\hline
		Task&Paper title&Skills&Authors\\ \hline
		\multirow{4}{*}{T1}&\multirow{3}{*}{\parbox{\linewidth}{A Unified Approach to Ranking  in Probabilistic Databases}} & 	\multirow{3}{*}{\parbox{\linewidth}{ approach, databases, probabilistic, ranking, unified}} &  Jian Li, \\ 
		& & & Barna Saha, \\ 
		& & & Amol Deshpande \\  \hline
		\multirow{4}{*}{T2}&	\multirow{4}{*}{\parbox{\linewidth}{Semantic Web Data Management Using Vertical Partitioning}} & 	\multirow{4}{*}{\parbox{\linewidth}{data, management, partitioning, scalable, semantic, using, vertical, web}} & Kate Hollenbach,  \\ 
		&  & & Samuel R. Madden, \\ 
		&  & & Adam Marcus, \\ 
		&  & & Daniel J. Abadi \\ \hline
		\multirow{4}{*}{T3}&	\multirow{4}{*}{\parbox{\linewidth}{Dense Subgraph Maintenance under Streaming Edge Weight Updates for Real-time Story Identification }} & 	\multirow{4}{*}{\parbox{\linewidth}{dense, edge, identification, maintenance, real, streaming, subgraph, time, updates, weight}} &  Albert Angel, \\ 
		&	&  & Nick Koudas, \\ 
		&	&  & Nikos Sarkas,\\ 
		&	&  & Divesh Srivastava \\ \hline
	\end{tabular}
	\caption{\textbf{Case study} : Three tasks T1, T2 and T3 are best papers in  VLDB (2009, 2007 and 2012 respectively) in the order of increasing skills }
	\label{case_tasks}
\end{table}


\begin{table}
	\centering
	\begin{tabular}{|l|l|l|l|}
		\hline
		Network&T1&T2&T3\\ \hline
		\multirow{3}{*}{DBLP}&\multirow{3}{*}{Haixun Wang}&\multirow{3}{*}{Samuel Madden}&Nick Koudas,\\
                &&&David Eppstein, \\
		&&&Dimitrios Gunopulos \\ \hline
		\multirow{5}{*}{DB}&Raymond T. Ng, &\multirow{2}{*}{Samuel Madden}&Divesh Srivastava \\
 	        &Jiawei Han&& Abraham Silberschatz, \\
                &&&Dimitrios Gunopulos, \\
		&&&Yehoshua Sagiv,\\
                &&&Lise Getoor \\ \hline
		\multirow{5}{*}{DM}&Yi Chang&Hui Xiong,&Jiawei Han,\\
                &Jiawei Han& Qiang Yang 0001& Qiaozhu Mei,\\
 	        &&&  C. Lee Giles,\\&&& George Karypis,\\&&& Wei Ding 0003\\ \hline
		\multirow{2}{*}{AI}	&Thomas S. Huang,&Gang Zeng,&Michael I. Jordan, Fei Sha, \\
                & Shuicheng Yan& Xian-Sheng Hua,& Alan S. Willsky,\\
		& Marc Pollefeys,&&Vittorio Murino, \\
                & Xindong Wu&&Larry S. Davis$ ^{*} $\\ \hline
		\multirow{3}{*}{TH}&Rajeev Motwani,&Amos Fiat,&S. Muthukrishnan,\\
                & Joseph Naor& Haim Kaplan,& David Eppstein, \\
                & Frank Mcsherry,&&Roberto Grossi,\\
                & Madhu Sudan&&Andris Ambainis\\ \hline
		\multirow{2}{*}{VLDB}&Surajit Chaudhuri,&Samuel Madden&\\
                & Raymond T. Ng&&\\\hline
		\multirow{3}{*}{SIGMOD}&Ruoming Jin,&Jeffrey F. Naughton,&\\
                & Jiawei Han &S. Sudarshan 0001,&\\
                &&Shalom Tsur&\\ \hline
		\multirow{5}{*}{ICDE}&Xuemin Lin,&Philip S. Yu,&Beng Chin Ooi, \\
                & Zhifeng Bao& Sanghyun Park& Kian-Lee Tan,Yueguo Chen,\\
		&&& Elke A. Rundensteiner, \\
		&&&Wynne Hsu, Yinghui Wu, \\
                &&&Stanley B. Zdonik$ ^{*} $\\ \hline
		\multirow{3}{*}{WWW}&C. Lee Giles, Hang Li,&Xin Li, Wei-Ying Ma,&\\
                & Pavel Serdyukov& Weiguo Fan, Xin Qi$^{*}$&\\
		&& Dongmei Wang$^{*}$,&\\
                && Alan Mislove$^{*}$&\\ \hline
		\multirow{2}{*}{KDD}&Jiawei Han,&Jian Pei, Haixun Wang,&\\
                & Anthony K. H. Tung,& Jiawei Han,&\\
		& Michael R. Lyu& Ke Wang, Bin Gao&\\ \hline
		\multirow{4}{*}{ICDM}&Zheng Chen,&&Philip S. Yu, Suh-Yin Lee,  \\ 
                &Philip S. Yu,&& Jeffrey Xu Yu \\
                & Jun Yan&&Bing Liu 0001, Bin Wu,\\
		&&&Wei Ding 0003, Qiaozhu Mei,\\ \hline
		\multirow{4}{*}{CVPR}&Thomas S. Huang,&Changshui Zhang, &\\
                & Jianchao Yang& Shipeng Li,&\\
                &&Nils Krahnstoever, &\\
		&&Marc Pollefeys&\\ \hline
		\multirow{5}{*}{STOC}&Mihalis Yannakakis,&&Venkatesan Guruswami, \\
                & Michael Sipser, && Sudipto Guha, \\
		&Jeffrey D. Ullman,& &Oded Regev,\\
                & Baruch Schieber&&Yury Makarychev, Brent Waters\\ 
		&&&Baruch Awerbuch, Moses Charikar, \\ \hline
	\end{tabular}
	\caption{Results of TFC-R algorithm for tasks T1, T2 and T3}
	\label{case_results}
\end{table}


\section{Conclusions and future work}	\label{sec:conc}

Generally for all the algorithms of team formation in the literature, the starting point is skill rather than an expert. The skills are ordered according to a greedy heuristic and the expert of the skill is chosen based on his/her  distance to the team already formed. 
In fact, in the case of Rarestfirst algorithm, the pre-processing step requires the distances between all the pairs of experts to be computed apriori. On large networks this step takes a few days even on systems with  reasonable configuration, hence the implementation has been modified by \cite{Wang-comparative-2015}.  
The core principle of our algorithm is to ensure that traversal of the entire network does not happen even in the worst case scenario. Our heuristic for choice of a leader uses nodes from the heavy tail ensured by the important property of power law underlying the degree distribution of a social network. Further we show that searching within a community is enough. We simply take members from 2 or 3 hops from the leader and the rest of the skill gap is filled by choosing a nearest neighbour or randomly. We tabulate in Figure~\ref{tfc_rand} to show the number of times this random step was needed. It shows that even for tasks of size 14,  70\% of the teams have not needed a random expert to be added in the case of DB and DBLP. 
TFC-N shows very good performance by managing the trade-off between time and cost efficiently.

We computed numerically the different ratios of the running time taken of our algorithms with respect to the others given in Figure~\ref{all-processing}. Our algorithms are 22.3   and 7.3 times faster when compared to MinSD  and MinLD respectively on average in finding a team for a given task.  In fact, TFC-R shows 95\% and 86\% improvement on average time taken when compared to MinSD and MinLD respectively. 

Further, in community-wise performance analysis seen in Section \ref{sec:results-comm}, it is interesting to see how fast the algorithms run on the smallest community of VLDB. The communication costs on VLDB are only slightly higher even though it is almost 1/23rd of the whole network. Further the communication costs obtained by the algorithms on DB community which is almost 1/5th of the size of DBLP network are as good as the results obtained on the whole network validating the scalability of our proposed approach. Of course it is still an question to be explored on  how to choose an \textquoteleft apt' community, so to say, for the team formation.

The case-study discussed in the paper shows the versatality of the proposed algorithms. They can produce multiple optimal teams for a task and it is part of our future work to evaluate their ability to produce teams with diversity.
It will be interesting to repeat the experimentation to datasets other than DBLP  like Bibsonomy, IMDB, Stackoverflow, github etc. which is part of future work. Also the many variations of team formation problem like capacitated TF, TF with personnel cost, TF  with budget constraints etc can be addressed with our community-based approach to team formation.

\bibliographystyle{abbrvnat}
\bibliography{tfc_draft}

\begin{thebibliography}{24}
\providecommand{\natexlab}[1]{#1}
\providecommand{\url}[1]{\texttt{#1}}
\expandafter\ifx\csname urlstyle\endcsname\relax
  \providecommand{\doi}[1]{doi: #1}\else
  \providecommand{\doi}{doi: \begingroup \urlstyle{rm}\Url}\fi

\bibitem[Anagnostopoulos et~al.(2010)Anagnostopoulos, Becchetti, Castillo,
  Gionis, and Leonardi]{Aris-power-2010}
A.~Anagnostopoulos, L.~Becchetti, C.~Castillo, A.~Gionis, and S.~Leonardi.
\newblock Power in unity: Forming teams in large-scale community systems.
\newblock In \emph{Proceedings of the 19th ACM International Conference on
  Information and Knowledge Management}, pages 599--608, 2010.

\bibitem[Anagnostopoulos et~al.(2012)Anagnostopoulos, Becchetti, Castillo,
  Gionis, and Leonardi]{Aris-online-2012}
A.~Anagnostopoulos, L.~Becchetti, C.~Castillo, A.~Gionis, and S.~Leonardi.
\newblock Online team formation in social networks.
\newblock In \emph{Proceedings of the 21st International Conference on World
  Wide Web}, pages 839--848, 2012.

\bibitem[Baghel and Bhavani(2018)]{Singh-multiple-2018}
V.~S. Baghel and S.~D. Bhavani.
\newblock {Multiple Team Formation Using an Evolutionary Approach}.
\newblock In \emph{2018 Eleventh International Conference on Contemporary
  Computing (IC3)}, pages 1--6. IEEE, 2018.

\bibitem[Demirovi{\'c} et~al.(2018)Demirovi{\'c}, Schwind, Okimoto, and
  Inoue]{Demirovic-recoverable-2018}
E.~Demirovi{\'c}, N.~Schwind, T.~Okimoto, and K.~Inoue.
\newblock Recoverable team formation: Building teams resilient to change.
\newblock In \emph{Proceedings of the 17th International Conference on
  Autonomous Agents and MultiAgent Systems}, pages 1362--1370, 2018.

\bibitem[Gajewar and Sarma(2012)]{Amita-multi-2012}
A.~Gajewar and A.~D. Sarma.
\newblock Multi-skill collaborative teams based on densest subgraphs.
\newblock In \emph{Proceedings of the 2012 SIAM International Conference on
  Data Mining}, pages 165--176, 2012.

\bibitem[Gaston and desJardins(2003)]{Gaston-complex-2003}
M.~Gaston and M.~desJardins.
\newblock Team formation in complex networks.
\newblock \emph{Proceedings of the 1st NAACSOS Conference}, 2003.

\bibitem[Gulla(2020)]{Joseph-2020}
J.~Gulla.
\newblock {7 Reasons IT Projects Fail}, Jul 2020.
\newblock URL
  \url{https://ibmsystemsmag.com/IBM-Z/02/2012/7-reasons-it-projects-fail}.
\newblock [Online; accessed 11. Jul. 2020].

\bibitem[Guti{\'e}rrez et~al.(2016)Guti{\'e}rrez, Astudillo,
  Ballesteros-P{\'e}rez, Mora-Meli{\`a}, and
  Candia-V{\'e}jar]{Gutierrez-multiple-2016}
J.~H. Guti{\'e}rrez, C.~A. Astudillo, P.~Ballesteros-P{\'e}rez,
  D.~Mora-Meli{\`a}, and A.~Candia-V{\'e}jar.
\newblock The multiple team formation problem using sociometry.
\newblock \emph{Computers \& Operations Research}, 75:\penalty0 150 -- 162,
  2016.

\bibitem[Huang(2020)]{Huang-2020}
J.~Huang.
\newblock {Best paper awards}, Jun 2020.
\newblock URL \url{https://jeffhuang.com/best_paper_awards.html}.
\newblock [Online; accessed 12. Jul. 2020].

\bibitem[Josh(2019)]{Josh-2019}
Josh.
\newblock {Experts Share Thoughts: A Team Collaboration Strategy That Helped
  Them Save Projects From Failing}, May 2019.
\newblock URL
  \url{https://www.proofhub.com/articles/team-collaboration-strategy}.
\newblock [Online; accessed 11. Jul. 2020].

\bibitem[Kargar and An(2011)]{Kargar-discovering-2011}
M.~Kargar and A.~An.
\newblock Discovering top-k teams of experts with/without a leader in social
  networks.
\newblock In \emph{Proceedings of the 20th ACM International Conference on
  Information and Knowledge Management}, pages 985--994, 2011.

\bibitem[Kargar et~al.(2012)Kargar, An, and Zihayat]{Kargar-efficient-2012}
M.~Kargar, A.~An, and M.~Zihayat.
\newblock Efficient bi-objective team formation in social networks.
\newblock In \emph{Machine Learning and Knowledge Discovery in Databases},
  pages 483--498, 2012.

\bibitem[Kargar et~al.(2013)Kargar, Zihayat, and An]{Kargar-affordable-2013}
M.~Kargar, M.~Zihayat, and A.~An.
\newblock Finding affordable and collaborative teams from a network of experts.
\newblock In \emph{Proceedings of the 2013 SIAM International Conference on
  Data Mining}, pages 587--595, 2013.

\bibitem[Lappas et~al.(2009)Lappas, Liu, and Terzi]{Lappas-finding-2009}
T.~Lappas, K.~Liu, and E.~Terzi.
\newblock Finding a team of experts in social networks.
\newblock In \emph{Proceedings of the 15th ACM SIGKDD International Conference
  on Knowledge Discovery and Data Mining}, pages 467--476, 2009.

\bibitem[{Li} and {Shan}(2010)]{Li-generalized-2010}
C.~{Li} and M.~{Shan}.
\newblock Team formation for generalized tasks in expertise social networks.
\newblock In \emph{2010 IEEE Second International Conference on Social
  Computing}, pages 9--16, 2010.

\bibitem[Li et~al.(2018)Li, Huang, and Yan]{Li-influence-2018}
C.-T. Li, M.-Y. Huang, and R.~Yan.
\newblock Team formation with influence maximization for influential event
  organization on social networks.
\newblock \emph{World Wide Web}, 21:\penalty0 939--959, 2018.

\bibitem[Li et~al.(2015)Li, Tong, Cao, Ehrlich, Lin, and
  Buchler]{Li-replacing-2015}
L.~Li, H.~Tong, N.~Cao, K.~Ehrlich, Y.-R. Lin, and N.~Buchler.
\newblock {Replacing the Irreplaceable: Fast Algorithms for Team Member
  Recommendation}.
\newblock In \emph{Proceedings of the 24th International Conference on World
  Wide Web}, pages 636--646. International World Wide Web Conferences Steering
  Committee, 2015.

\bibitem[Majumder et~al.(2012)Majumder, Datta, and
  Naidu]{Majumder-capacited-2012}
A.~Majumder, S.~Datta, and K.~Naidu.
\newblock Capacitated team formation problem on social networks.
\newblock In \emph{Proceedings of the 18th ACM SIGKDD International Conference
  on Knowledge Discovery and Data Mining}, pages 1005--1013, 2012.

\bibitem[Marr(2016)]{Marr-2016}
B.~Marr.
\newblock {Are These The 7 Real Reasons Why Tech Projects Fail?}, Sep 2016.
\newblock URL
  \url{https://www.forbes.com/sites/bernardmarr/2016/09/13/are-these-the-real-reasons-why-tech-projects-fail/#5ae78be67320}.

\bibitem[McDonald(2003)]{McDonald-recommend-2003}
D.~W. McDonald.
\newblock Recommending collaboration with social networks: A comparative
  evaluation.
\newblock In \emph{Proceedings of the SIGCHI Conference on Human Factors in
  Computing Systems}, pages 593--600, 2003.

\bibitem[Newman(2006)]{Newman-modularity-2006}
M.~E.~J. Newman.
\newblock {Modularity and community structure in networks}.
\newblock \emph{Proc. Natl. Acad. Sci. U.S.A.}, 103:\penalty0 8577--8582, 2006.

\bibitem[Tavrizyan(2019)]{Tavrizyan-2019}
K.~Tavrizyan.
\newblock {16 Project Management Stats You Can{'}t Ignore [2019]}, Jun 2019.
\newblock URL
  \url{https://medium.com/crowdbotics/hips-dont-lie-15-project-management-stats-you-can-t-ignore-6f655060ef30}.
\newblock [Online; accessed 11. Jul. 2020].

\bibitem[Wang et~al.(2015)Wang, Zhao, and Ng]{Wang-comparative-2015}
X.~Wang, Z.~Zhao, and W.~Ng.
\newblock A comparative study of team formation in social networks.
\newblock In \emph{Database Systems for Advanced Applications}, pages 389--404,
  2015.

\bibitem[Wi et~al.(2009)Wi, Oh, Mun, and Jung]{Wi-knowledge-2009}
H.~Wi, S.~Oh, J.~Mun, and M.~Jung.
\newblock A team formation model based on knowledge and collaboration.
\newblock \emph{Expert Syst. Appl.}, 36:\penalty0 9121--9134, 2009.

\end{thebibliography}

\appendix\label{appendix}

\setcounter{table}{0}

\section{Networks}

\begin{table}
	\centering
	\begin{tabular}{|l|l|l|l|l|l|}
		\hline
		Network&$|V|$&$|E|$&$|S|$&$d_{avg}$&$\frac{|S|}{|E|}$\\ \hline
		DBLP&32477&98676&13232&6.08&17.24\\ \hline
		DB&6053&19607&4310&6.48&14.58\\ \hline
		DM&8323&23667&4912&5.69&13.93\\ \hline
		AI&14197&36514&7605&5.14&15.62\\ \hline
		TH&5401&16399&5626&6.07&19.94\\ \hline
		VLDB&1391&3605&1609&5.18&11.30\\ \hline
		SIGMOD&2020&5865&1948&5.81&11.29\\ \hline
		ICDT&190&402&461&4.23&11.80\\ \hline
		ICDE&2977&8392&2406&5.64&11.48\\ \hline
		PODS&549&1203&1029&4.38&12.03\\ \hline
		WWW&1904&4827&1788&5.07&10.40\\ \hline
		SDM&642&1605&898&5.00&11.37\\ \hline
		KDD&2111&5906&2036&5.60&11.54\\ \hline
		ICDM&1945&4894&1897&5.03&12.04\\ \hline
		PKDD&220&414&499&3.76&10.66\\ \hline
		WSDM&378&944&540&4.99&9.89\\ \hline
		NIPS&3887&8647&3465&4.45&13.26\\ \hline
		IJCAI&2448&4878&2814&3.99&11.28\\ \hline
		ICML&2083&4229&2039&4.06&11.69\\ \hline
		UAI&794&1459&1408&3.68&12.80\\ \hline
		COLT&446&903&1015&4.05&14.00\\ \hline
		CVPR&5291&14351&3534&5.42&15.38\\ \hline
		FOCS&1425&3444&2297&4.83&15.20\\ \hline
		SODA&1883&4665&2418&4.95&13.56\\ \hline
		STOC&1349&3328&2267&4.93&16.26\\ \hline
		ICALP&1418&2751&2109&3.88&11.34\\ \hline
		STACS&435&781&946&3.59&10.81\\ \hline
		ESA&865&1885&1240&4.36&10.35\\ \hline
	\end{tabular}
	\caption{Network statistics}
	\label{ns}
\end{table}

\end{document}